\PassOptionsToPackage{svgnames}{xcolor}
\documentclass[twocolumn]{aastex631}
\usepackage{fixfoot}
\usepackage[clockwise, figuresright]{rotating}
\usepackage{amsmath}
\usepackage{commath}
\usepackage{gensymb}
\usepackage{color,soul}
\usepackage{multirow}
\usepackage{IEEEtrantools}
\usepackage{bm}
\usepackage{float}
\usepackage{textcomp}
\usepackage{booktabs}
\usepackage{csquotes}
\usepackage{morefloats}
\usepackage{scalerel}
\usepackage{graphicx,array}

\usepackage{appendix}
\defcitealias{Prugniel1997}{PS}
\submitjournal{ApJ}

\shorttitle{ $M_{\rm BH}$--Host Spheroid Density}
\shortauthors{Sahu, Graham, and Davis}

\begin{document}


\title{The (Black Hole Mass)-(Spheroid Stellar Density) Relations: $M_{\rm BH}$--$\mu$ (and $M_{\rm BH}$--$\Sigma$) and $M_{\rm BH}$--$\rho$}

\correspondingauthor{Nandini Sahu}
\email{nsahu@swin.edu.au}

\author[0000-0003-0234-6585]{Nandini Sahu}
\affil{OzGrav-Swinburne, Centre for Astrophysics and Supercomputing, Swinburne University of Technology, Hawthorn, VIC 3122, Australia}
\affil{Centre for Astrophysics and Supercomputing, Swinburne University of Technology, Hawthorn, VIC 3122, Australia}
\affil{Currently hosted by the University of Queensland, Brisbane,  QLD 4067, Australia}

\author[0000-0002-6496-9414]{Alister W. Graham}
\affil{Centre for Astrophysics and Supercomputing, Swinburne University of Technology, Hawthorn, VIC 3122, Australia}

\author[0000-0002-4306-5950]{Benjamin L. Davis}
\affil{Center for Astro, Particle, and Planetary Physics (CAP$^3$), New York University Abu Dhabi}



\begin{abstract}

This paper is the fourth in a series presenting  (galaxy morphology, and thus galaxy formation)-dependent black hole mass, $M_{\rm BH}$, scaling relations.  We have used a sample of 119  galaxies with directly-measured $M_{\rm BH}$ and host spheroid parameters obtained from multi-component decomposition of, primarily, $3.6\,\mu$m Spitzer images. Here, we investigate the correlations between $M_{\rm BH}$ and the projected luminosity density $\mu$, the projected stellar mass density $\Sigma$, and the deprojected (internal) stellar mass density $\rho$, for various spheroid radii. 
We discover the predicted $M_{\rm BH}$--$\mu_{\rm 0,sph}$ relation and present the first $M_{\rm BH}$--$\mu_{\rm e, sph}$ and $M_{\rm BH}$--$\rho_{\rm e,int, sph}$  diagrams displaying slightly different (possibly curved) trends for early- and late-type galaxies (ETGs and LTGs) and an offset between ETGs with (fast-rotators, ES/S0) and without (slow-rotators, E) a disk. 
The scatter about various $M_{\rm  BH}$--$\langle\Sigma\rangle_{\rm R,sph}$ (and $\langle\rho\rangle_{\rm r,sph}$) relations is shown to systematically decrease as the enclosing aperture (and volume) increases, dropping from 0.69~dex when using the spheroid  \enquote{compactness}, $\langle\Sigma\rangle_{\rm 1kpc,sph}$, to 0.59~dex when using $\langle\Sigma\rangle_{\rm 5kpc,sph}$. 
We also reveal that $M_{\rm BH}$ correlates with the internal density,  $\rho_{\rm soi,sph}$, at the BH's sphere-of-influence radius, such that  core-S\'ersic (high S\'ersic index, $n$) and (low-$n$) S\'ersic galaxies define different relations with total rms scatters 0.21~dex and 0.77~dex, respectively. 
The $M_{\rm BH}$--$\langle\rho\rangle_{\rm soi,sph}$ relations shall help with direct estimation of tidal disruption event rates, binary BH lifetimes, and together with other BH scaling relations, improve the characteristic strain estimates for long-wavelength gravitational waves pursued with pulsar timing arrays and space-based interferometers.



\end{abstract}

\section{Introduction}

The number of galaxies with directly-measured black hole masses, i.e., where observations could resolve the black hole's gravitational sphere-of-influence, has grown to about 145 galaxies \citep{Sahu:2019:b}. 
Using state-of-the-art two-dimensional modeling \citep{Ciambur:2015:Ellipse} and multi-component decompositions \citep{Ciambur:2016:Profiler},  we have modeled the surface brightness profiles of 123 of these galaxies\footnote{This was the known sample size in 2018 when this projected commenced.} and their components.
We have discovered  morphology-dependent correlations between the black hole mass ($M_{\rm BH}$) and various host galaxy properties, such as the  galaxy stellar mass ($M_{\rm *, gal}$), the spheroid stellar mass ($M_{\rm *,sph}$),  the spheroid  central light concentration or S\'ersic index ($n_{\rm sph}$), the spheroid effective half-light radius ($R_{\rm e,sph}$), and the central stellar velocity dispersion \citep{Graham:2012, Graham:Scott:2013, Scott:Graham:2013, Savorgnan:2016:Slopes, Davis:2018:b, Davis:2018:a, Sahu:2019:a, Sahu:2019:b, Sahu:2020}. 
These have improved as the quality, and the quantity of data has grown.
The simple  (galaxy morphology)-independent black hole scaling relations\footnote{The history of the galaxy/black hole scaling relations is reviewed in \citet{Ferrarese:Ford:2005} and \citet{Graham:2016:Review}.}  \citep[e.g.,][]{Dressler:Richstone:1988, Magorrian:1998, Haring:Rix:2004, Gultekin:2009, Kormendy:Ho:2013, McConnell:Ma:2013} are, in fact, too simple to accurately trace the coevolution of the different types of galaxies and their black holes. 
For example, the small and massive bulges of spiral and lenticular galaxies follow $M_{\rm BH}$--$M_{\rm *,sph}$ relations different from that of elliptical galaxies \citep{Sahu:2019:a}. 
It is hoped that the advances with morphology-dependent correlations will help identify which correlation is more fundamental, i.e., primary versus secondary. However, one of the potential candidates remains to be explored; it involves stellar density.

Most of the morphology-dependent black hole scaling relations are significantly different to the familiar but now superseded \enquote{single relations} obtained when all galaxy types are combined.   Crucially, diagrams with differing numbers of different galaxy types can yield  \enquote{single relations} with different slopes and intercepts.  As a result, many of the past \enquote{single relations} (built by grouping galaxies of different morphological types) are physically meaningless, merely representative of the ratio of the different galaxy types in that sample. 

The above realization is fundamental if we are to adequately understand the co-evolution of galaxies and their central massive black holes. This is because the black hole mass is, in a sense, aware of the different formation history and physics which went into building its host galaxy. What is important is not simply the amount of mass in stars and perhaps dark matter, but how that mass was assembled (and moves) to create a galaxy's substructure/morphology.

Using a sample of 27 galaxies, \citet{Graham:Driver:2007}  observed a strong correlation between $M_{\rm BH}$  and the bulge central concentration, which is quantified by the shape parameter of spheroid's surface brightness profile the S\'ersic index  \citep[$n_{\rm sph}$,][]{Trujillo:Graham:Caon:2001}. 
\citet{Graham:Driver:2007}  found a comparable level of (intrinsic) scatter about the $M_{\rm BH}$--$n_{\rm sph}$ relation as seen in the $M_{\rm BH}$--(central stellar velocity dispersion: $\sigma$) relations observed at that time, which was about 0.3 dex. 
The observed stellar velocity dispersion traces the underlying mass distribution and radial concentration of light  \citep{Graham:Trujillo:Caon:2001}. 
Thus, \citet{Graham:Driver:2007}  suggested that a combination of the central stellar density and the central light concentration of a spheroid may be governing the black hole--host spheroid connection. 

Some studies \citep[e.g.,][]{Graham:Guzman:2003, Merritt:2006}  presented a  correlation between central concentration and central stellar density, suggesting that one of these quantities can be written in terms of the other; although, it is still not clear which quantity is more fundamental.  
\citet{Graham:Driver:2007} combined this relation with their  linear and curved\footnote{\citet{Graham:Driver:2007} also presented an even stronger but curved $M_{\rm BH}$--$n_{\rm sph}$ relation with an intrinsic scatter of 0.18 dex.}  $M_{\rm BH}$--$n_{\rm sph}$ relations, to predict  both a linear and a curved $M_{\rm BH}$--(spheroid central surface brightness or central projected density, $\mu_{\rm 0,sph}$) relation \citep[][their equations 9 and 10]{Graham:Driver:2007}.
Moreover, they suggested that an even better correlation might exist between  $M_{\rm BH}$ and the (three-dimensional) deprojected density ($\rho$, \textit{aka} the internal or spatial density) at the center of the spheroid.  
For the first time, here we explore these predicted $M_{\rm BH}$--(stellar density) relations and others.



We present new correlations between $M_{\rm BH}$ and the spheroid surface brightness (projected/column  luminosity density),  the projected (or column) stellar mass density ($\Sigma$), and the deprojected stellar density at various spheroid radii. 
Our sample of  123 galaxies is described in the following Section \ref{Data}. 
That section also describes the linear regression applied and the parameter uncertainty used. 
The calculation of the deprojected density is detailed in the Appendix \ref{calculation_density}, where we also compare our numerically calculated internal density with an approximation from the model of \citet{Prugniel1997}. 

Section \ref{Projected_density} presents the correlation between $M_{\rm BH}$ and the spheroid projected (luminosity and stellar mass)  density at various radii (center, 1 kpc, 5 kpc, and half-light radius). 
In Section \ref{Deprojected_density},  we reveal additional new correlations obtained between $M_{\rm BH}$ and the bulge internal mass density at various (inner and larger) radii, including the sphere-of-influence radius of the black hole. 
We also provide the projected and deprojected density profiles, $\mu(R)$ and  $\rho(r)$, to help explain various trends obtained between $M_{\rm BH}$ and $\mu$, and between $M_{\rm BH}$ and $\rho$ at different spheroid radii.

In all of these diagrams, we also investigate possible dependence on galaxy morphology, e.g., early-type galaxies (ETGs: elliptical E, ellicular ES\footnote{Ellicular galaxies have an intermediate-scale, rotating stellar disk fully confined within their bulge \citep{Liller:1966, Savorgnan:Graham:ES:2016}. The term ellicular is a concatenation made by combining the words \enquote{elliptical} and \enquote{lenticular}  \citep[see][ for a historical review of galaxy morphology and classification schemes]{Graham:Grid:2019}.}, and lenticular S0) versus late-type galaxies (LTGs: spirals S), centrally-fast (ES, S0, S) versus slow (E) rotators,  core-S\'ersic\footnote{The core-S\'ersic galaxies are generally the most massive galaxies, likely formed through major gas-poor mergers. The eventual coalescence of their central massive black holes scours out the stars from the central \enquote{loss cone} (through the transfer of the binary black hole's orbital angular momenta) and creates a deficit of light at the center, referred to as a \enquote{core}. The bulge surface brightness profile for a core-S\'ersic  galaxy is described by a core-S\'ersic function \citep{Graham:2003:CS}, which consists of a shallow inner power-law followed by a S\'ersic function \citep{Sersic:1968, Sersic:1963} at larger radii. Such cores were first noted by \citet{King:Minkowski:1966}.} versus S\'ersic\footnote{S\'ersic galaxies do not have a deficit of light at their center.}  galaxies, and barred versus non-barred galaxies. 
We compare our findings with the morphology-dependent substructures seen in our recently published correlations \citep{Sahu:2019:a, Sahu:2019:b, Sahu:2020}. In Section \ref{Discussion} we discuss our results and some of the more notable implications. Finally,  we summarize the main results of this work in  Section \ref{conclusions}. 

We have used the terms spatial density, and internal density interchangeably for the (3D) deprojected density throughout this paper.  All uncertainties are quoted at the $\pm 1\,\sigma \, (\approx 68\%)$ confidence interval.

\section{Data} 
\label{Data}

We have used the spheroid's structural parameters from \citet{Savorgnan:Graham:2016:I}, \citet{Davis:2018:a}, and \citet{Sahu:2019:a}, which were obtained from multi-component decompositions of 123 galaxies with directly-measured central black hole masses reported in the literature. 
The direct methods for black hole mass measurement include stellar dynamical modeling, gas dynamical modeling, megamaser kinematics, proper motions (for Sgr $A^*$), and the latest direct imaging (for M87*). The majority of the galaxy images ($\rm 81 \%$)  were in the $3.6\,\mu$m-band taken by the infrared array camera 
 \citep[IRAC, ][resolution $\sim 2 \arcsec$]{Fazio:Hora:2004} onboard the Spitzer Space Telescope. 
The remaining images came from the archives of the Hubble Space Telescope (HST, $11 \%$), the Sloan Digital Sky Survey (SDSS, $3 \%$), and the Two Micron All Sky Survey (2MASS, $6 \%$). 
For full details of the image analysis, we refer readers to the aforementioned three studies.

Briefly, we performed 2D modeling of the galaxy images using our in-house\footnote{The software \textsc{isofit}, \textsc{cmodel},  and \textsc{profiler} are publicly available at the GitHub platform \citep[see][for details]{Ciambur:2015:Ellipse, Ciambur:2016:Profiler}.} software \textsc{isofit} and \textsc{cmodel} \citep{Ciambur:2015:Ellipse}, which were built into the image reduction and analysis facility \citep[\textsc{iraf},][]{Tody:1986, IRAF}.  \textsc{isofit} fits quasi-elliptical isophotes at each galactic radii.  
It uses an elliptical coordinate system, thereby improves upon the spherical coordinate system implemented in \textsc{ellipse} \citep{Jedrzejewski1:1987, Jedrzejewski2:1987}. 
The angular coordinate known as the  \enquote{eccentric anomaly} is used for uniform sampling of the quasi-elliptical isophotes, and the code employs Fourier harmonics to capture the isophotal deviations from a pure ellipse \citep{Carter:1978, Kent:1984, Michard:Simien:1988}. 
Thus,  \textsc{isofit} generates an (azimuthally-averaged) one-dimensional surface brightness profile along any galaxy axis, together with the radial variations of the isophotal ellipticity ($e$), position angle, and Fourier coefficients. 
These parameters are used to create a 2D  galaxy model via \textsc{cmodel}. 
The model captures all symmetric features about the major-axis (mirror symmetry) and leaves behind disturbances and star clusters which can be explored in the \enquote{residual image}.

We disassemble the galaxy model into components with the help of various functions inbuilt in the software \textsc{profiler} \citep{Ciambur:2016:Profiler}. 
A galaxy can have a bulge, intermediate- or large-scale disk, bar, ansae, rings,  depleted core, and nuclear components (e.g., star cluster, nuclear bar, disk, or ring). 
The presence of disks and bars in our decompositions were verified, whenever possible, through recourse to the literature, including kinematic evidence for disk rotation. 
We perform this multi-component decomposition using the surface brightness profile along the galaxy's major-axis as well as the so-called \enquote{equivalent-axis}, which represents a radial-axis equivalent to a circularised form of the galaxy's quasi-elliptical isophotes, such that the total enclosed luminosity remains conserved\footnote{The equivalent-axis radii, $R_{\rm eq}$, for an isophote is the geometric-mean of the isophote's  major- and minor-axis radii ($R_{\rm maj}$ and $R_{\rm min}$, respectively), i.e., $R_{\rm eq}=\sqrt{R_{\rm maj} \times R_{\rm min}}$ or $ R_{\rm eq}=R_{\rm maj} \sqrt{1-e_{\rm maj}}$ \citep[see][for more details on the isophotal galaxy modeling, multi-component decomposition, and the circularised equivalent-axis]{Ciambur:2015:Ellipse, Ciambur:2016:Profiler}.}.

\begin{figure*}
\begin{center}
\includegraphics[clip=true,trim= 06mm 00mm 00mm 13mm,width= 1\textwidth]{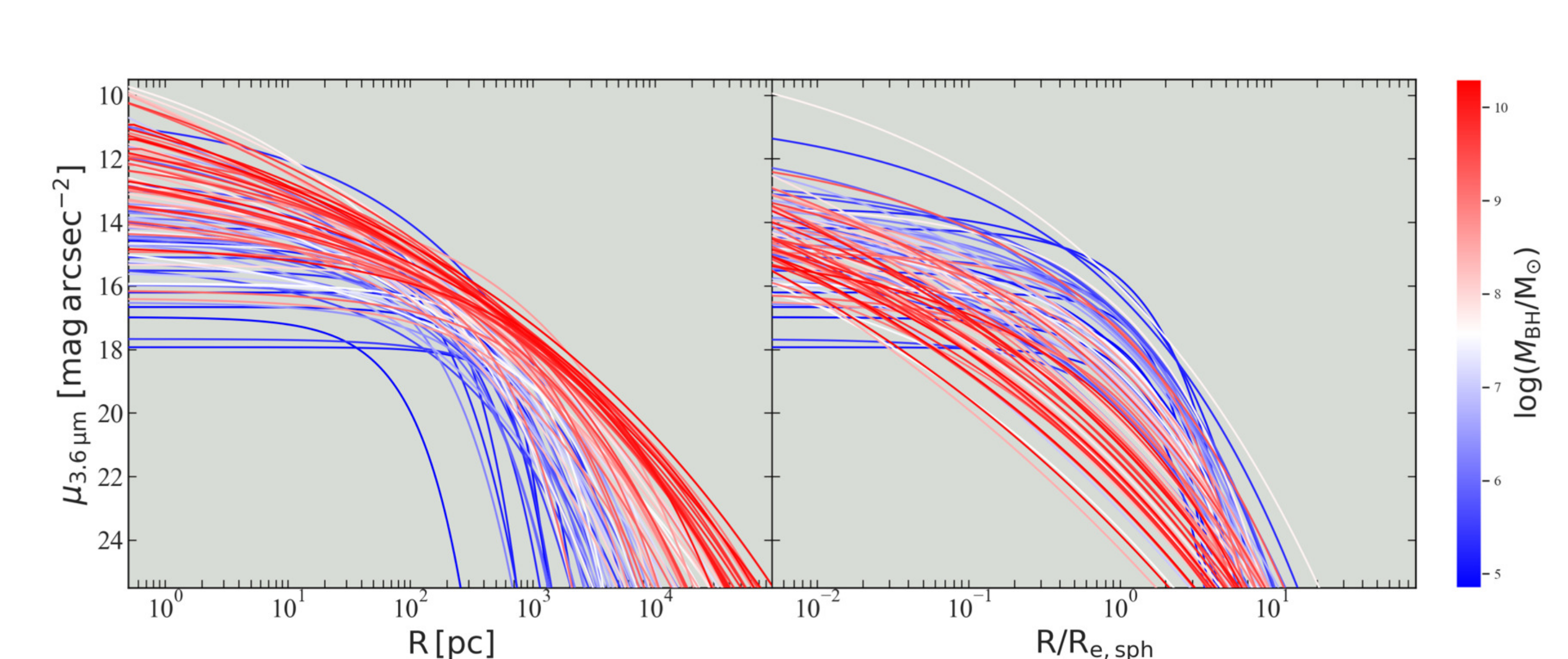}
\caption{Left-hand panel: Spheroid (S\'ersic) surface brightness profiles for our $\rm 3.6\, \mu$m-sample. Right-hand panel: The horizontal axis is normalized at the (projected) half-light radii of each spheroid. The color sequence blue-white-red traces the increasing black hole mass and helps with the understanding of the positive/negative trends observed between $M_{\rm BH}$ and the spheroid surface brightness (and the projected stellar mass density) presented in Section \ref{Projected_density}.
}
\label{Sersic_all}
\end{center}
\end{figure*}

The multi-component decomposition process provides us with the surface brightness profiles of individual galaxy components and the detailed galaxy morphology, indicating the presence of a rotating disk, a depleted core, a bar, etc. One of the most noted of all galactic components is the bulge, whose surface brightness distribution is described using the \citet{Sersic:1963} function  (Appendix Equation \ref{Sersic}), which is parameterized by the S\'ersic index ($n_{\rm sph}$), effective half-light radius ($R_{\rm e, sph}$), and the surface brightness at the half-light radius ($\mu_{\rm e, sph}= -2.5\log I_{\rm e,sph}$). The equivalent-axis spheroid surface brightness profiles for our $\rm 3.6\,\mu m$ (Spitzer) sample are shown in Figure \ref{Sersic_all}.

To obtain the spheroid's internal (deprojected) stellar mass density distribution, $\rho(r)$, we performed an inverse Abel transformation \citep{Abel1826} of the (circularly symmetric) equivalent-axis spheroid surface brightness profiles. The numerical calculation of the internal density profiles and a comparison with the  \citet{Prugniel1997} density model (an approximation to the exact deprojection of the S\'ersic profile) is presented in the Appendix Section \ref{calculation_density}.

The spheroid parameters required to calculate the projected and the internal stellar mass densities, e.g., the bulge surface brightness parameters ($n_{\rm sph}, \, R_{\rm e, sph}, \, \mu_{\rm e, sph}$), along with the galaxy morphology, distances, physical (arcsec-to-kpc) scale\footnote{The arcsec-to-pc scale was calculated using cosmological parameters from \citet{Planck:Collaboration:2018}.}, stellar mass-to-light ratio, and the image band information for all 123 galaxies are available in \citet[][their Appendix Table A1]{Sahu:2020}. \citet{Sahu:2020} also tabulates the directly-measured central black hole masses and the bulge stellar masses ($M_{\rm *, sph}$) of these 123 galaxies.

Here, we excluded the galaxies  NGC~404, NGC~4342, NGC~4486B, and the Milky Way throughout our investigation, unless expressly stated otherwise. NGC~404  is the only galaxy with a black hole mass below $ 10^6 \, \rm M_\odot$ \citep{Nguyen:2017} and it may skew/bias the results. 
Its published black hole mass has a sphere of influence five times smaller than the seeing ($\rm \sim \, 0.1\arcsec$) under which it was measured. 
 NGC~4342 and NGC~4486B  have been heavily stripped of their mass due to the gravitational pull of their massive companion galaxies \citep[see][]{Batcheldor:2010:b, Blom:Forbes:2014}. 
For the Milky Way, the available surface brightness profile \citep{Kent:Dame:1991, Graham:Driver:2007} was not flux-calibrated to obtain a calibrated density profile. 
The exclusion of these galaxies leaves us with a reduced sample of 119 galaxies.  
All the galaxies excluded from the linear regressions (performed to obtain the scaling relations presented here) are shown with a different symbol in the ensuing diagrams.

We use the bivariate correlated errors and intrinsic scatter (\textsc{bces}) regression \citep{Akritas:Bershady:1996} to obtain our black hole scaling relations.   
\textsc{bces} is a modification of the ordinary least squares regression. It considers measurement errors in both variables (and their possible correlation) and allows for intrinsic scatter in the distribution. 
We prefer to use the \textsc{bces(bisector)}\footnote{The Python module written by \citep{Nemmen:2012} is available at \url{https://github.com/rsnemmen/BCES}.} line obtained by symmetrically bisecting the \textsc{bces($Y|X$)} line (which minimizes the error-weighted root mean square, rms, vertical offsets about the fitted line) and the  \textsc{bces($X|Y$)} line (which minimizes the error-weighted rms horizontal offsets about this  different fitted line). 
We do this partly because it is unknown whether the host spheroid density is an independent variable and the central black hole mass is a dependent variable, or vice-versa, or if there is an interplay.  
We also check our best-fit parameters using a symmetric application \citep{Novak:2006} of the intrinsically non-symmetric (modified \textsc{fitexy}) known as \textsc{mpfitexy}\footnote{Available at \url{https://github.com/mikepqr/mpfitexy}.} regression \citep{Markwardt:2009, Williams:2010}.

Uncertainties in $M_{\rm BH}$, and spheroid profile parameters $n_{\rm sph}$ ($\pm 0.09$ dex), $R_{\rm e,sph}$ ($\pm 0.13$ dex), and $\mu_{\rm e,sph}$ ($\pm 0.58 \, \rm mag \, arcsec^{-2}$ or $\pm 0.23$ dex in $\mu_{\rm e,sph}/2.5$) are taken from \citet[][see their  section 2 for more details]{Sahu:2020}. 
The uncertainty in the internal density ($\rho_e$) at an internal radius equal to the projected half-light radius ($R_{\rm e,sph}$)---obtained by propagating errors in the spheroid parameters through the analytical expression (Equation \ref{Rhoe_Ie})---are $\sim \pm 0.30$ dex. For densities at other radii, the error propagation (assuming independent parameters) through the internal density expression (Equation \ref{PS}) provides even higher uncertainties due to multiple occurrences of $n_{\rm sph}$ and $R_{\rm e,sph}$, in addition to $\rho_e$. 
Such uncertainties are likely to be overestimated and can affect the best-fit lines. 
Therefore, we used a constant uncertainty of $\pm 0.23$ dex on the projected mass densities ($\Sigma$) and, similarly, a constant uncertainty of $\pm 0.30$ dex on the internal densities for all the correlations, unless stated otherwise.  
Additionally, we test the stability of our correlations (their slopes and intercepts) using a range of (zero to $\pm 0.38$ dex) uncertainties for the projected and internal densities.

\section{Black Hole Mass versus Spheroid Projected Density}
\label{Projected_density}

\subsection{Central Surface Brightness and Projected Mass Density: $\rm \mu_0 \, \&  \, \Sigma_0$ }
\label{3.1}

\begin{figure*}
\begin{center}
\includegraphics[clip=true,trim= 9mm 03mm 15mm 13mm,width= 0.99\textwidth]{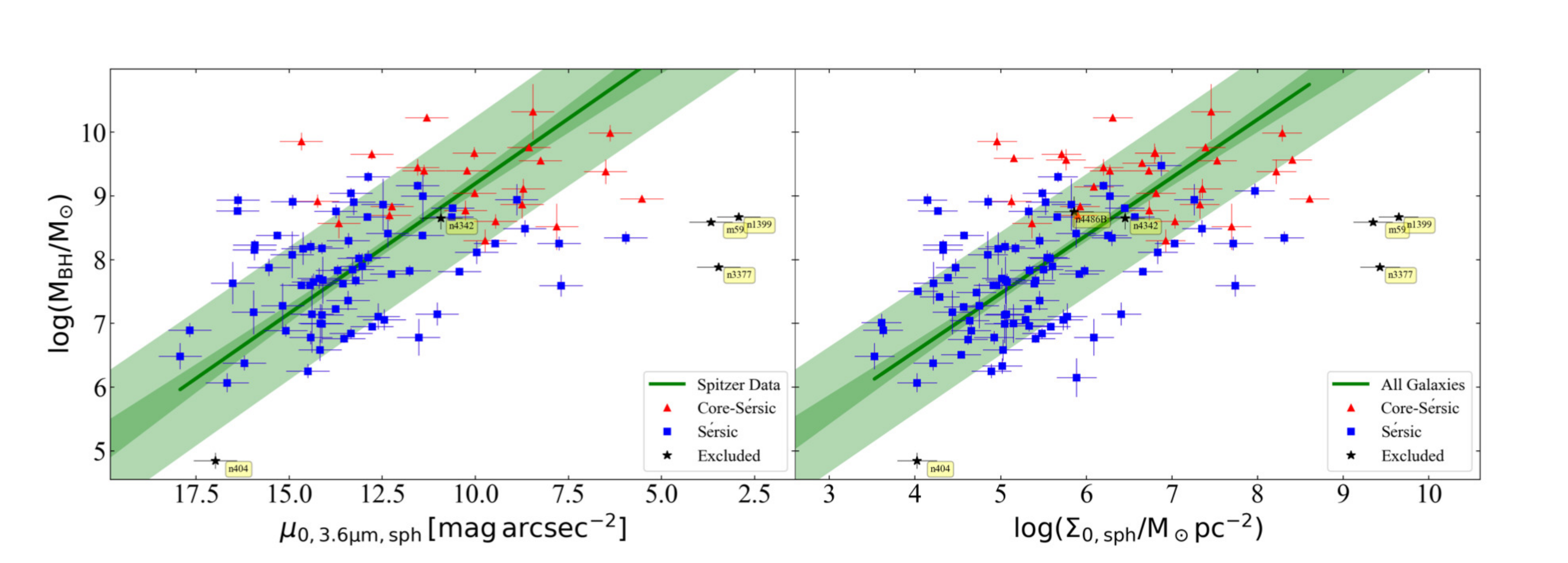}
\caption{Left-hand panel: Black hole mass versus the central surface brightness (in the AB magnitude system) of the spheroids using our $\rm 3.6 \, \mu m$ sample. Right-hand panel: Black hole mass versus the projected central stellar mass density, including the 22 non-Spitzer galaxies. The dark green line represents the best-fit obtained from the \textsc{bces(bisector)} regression. The dark green shaded region around the best-fit line delineates the $\pm 1\sigma$ uncertainty on the slope and intercept, and the light green shaded area outlines the $\pm 1 \sigma$ rms scatter in the data. The same description follows for all other correlations presented in this paper. For both S\'ersic and core-S\'ersic galaxies,  $\mu_{\rm 0, 3.6\mu m, sph}$ and $\Sigma_{\rm 0, sph}$ have been obtained through the inward extrapolation of the S\'ersic portion of their spheroid profiles. Core-S\'ersic galaxies have a deficit of light at their core and, hence, their $\mu_{\rm 0, 3.6\mu m, sph}$ values depicted here are brighter than the actual value. The galaxies excluded from the regression are marked with a black star.
}
\label{Mbh_I0_Ie}
\end{center}
\end{figure*}

To study the correlation between black hole mass and the host spheroid's central surface brightness ($\mu_{\rm 0, sph}$), which is dependent on the image wavelength band, we used our  $\rm 3.6\, \mu m$-sample comprised of 97 galaxies from the reduced sample of 119 galaxies (see Section \ref{Data}). This includes 72 S\'ersic galaxies, i.e., galaxies with a S\'ersic  spheroid surface brightness profile, and 25 core-S\'ersic galaxies, i.e., galaxies with a depleted central core whose spheroid profile is described by a shallow central power-law followed by a S\'ersic function at larger radii \citep[see][]{Graham:2003:CS}.

Using the (equivalent-axis)  surface brightness parameters  ($n_{\rm sph}$, $\rm R_{\rm e,sph}$, and  $\rm \mu_{\rm e,sph}$) for the spheroids, we calculated $\mu_{\rm 0, sph}$ via $\rm \mu_0=\mu_e -2.5\log e^{b}$ (Equation \ref{Sersic} at R=0), i.e., an inward extrapolation of the S\'ersic fit to the spheroid's surface brightness profile. 
It is important to note that for our core-S\'ersic galaxies, $\mu_0$ has been obtained through the inward extrapolation of the S\'ersic part of their spheroid profile \citep[as in the $L_{\rm gal}$--$\mu_0$ diagram of][]{Jerjen2000}. 
This is because the size of the depleted core is generally much smaller than the $\rm \sim \, 2\arcsec$ spatial resolution of IRAC images \citep[see][]{Dullo:2014}, and, as such, the ($\rm 3.6\, \mu m$-band) parameters for the central power-law of our core-S\'ersic spheroids are not accurate\footnote{The presence of the cores were confirmed through smaller field-of-view high-resolution HST images and the literature when available.}.  
Thus, the  $\mu_0$ used here for the cored galaxies represents their central surface brightness before the damaging effect of binary black holes, which will cause a departure of cored galaxies from an initial $M_{\rm BH}$--$\mu_0$  trend line. 
This is the case with cored ETGs in the $M_{\rm *,gal}$--$\mu_0$ diagram shown in \citet[][their figure 9]{Graham:Guzman:2003}, which accounted for the central mass/light deficit in the cored galaxies.


The high-$n_{\rm sph}$ galaxies M~59, NGC~1399 (cored), and NGC~3377 are marked by black stars and have the  brightest $\mu_{\rm 0, 3.6\mu m, sph}$ ($\sim 3 \rm \, mag \, arcsec^{-2}$) in  Figure \ref{Mbh_I0_Ie}. They reside beyond the $2\sigma$ scatter of the remaining dataset and have significant leverage on the best-fit line, such that including these three galaxies in the regression changes the slope by $1\sigma$. Therefore, these three galaxies were excluded from the regression (in addition to the four exclusions mentioned in Section \ref{Data}) to obtain the $M_{\rm BH}$--$\mu_{\rm 0, 3.6\mu m, sph}$ and the $M_{\rm BH}$--$\Sigma_{\rm 0,sph}$ relations reported here. 

The $M_{\rm BH}$--$\mu_{\rm 0, 3.6\mu m, sph}$ relation\footnote{The uncertainty we assigned to  $\mu_{\rm 0, 3.6\mu m, sph}$ is 0.58 $\rm mag \, arcsec^{-2}$ (the same as assigned to $\mu_{\rm e, 3.6\mu m, sph}$); however, consistent relations are obtained upon using up to 2 $\rm mag \, arcsec^{-2}$ uncertainty.}  plotted in the left-hand panel of Figure \ref{Mbh_I0_Ie} was obtained using 94 (S\'ersic +core-S\'ersic)  galaxies with $\rm 3.6\, \mu m$ imaging data, and can be expressed as,
\begin{IEEEeqnarray}{rCl}
\label{Mbh_mu0}
\rm  \log \left( \frac{M_{\rm BH}}{\rm M_\odot } \right)&=& \rm (-0.41 \pm 0.04)\, \left[\mu_{\rm 0, 3.6 \mu m, sph} -  \frac{13\, mag}{arcsec^{2}} \right]  \nonumber \\
&& +\> (7.97 \pm 0.10). 
\end{IEEEeqnarray} 
The total (measurement error and intrinsic scatter) rms scatter ($\Delta_{\rm rms|BH}$) is 1.03 dex in the $\log (M_{\rm BH})$-direction. This correlation quantifies how the (S\'ersic) spheroids hosting more massive black holes have a brighter central surface brightness, qualitatively consistent with the linear prediction of the $\log (M_{\rm BH})$--$\mu_{\rm 0, sph}$ relation in \citet[][their equation 9 based on B-band data]{Graham:Driver:2007}. This trend can also be inferred from the spheroid profiles plotted in the left-hand panel of Figure \ref{Sersic_all}, where for R tending to zero, $\mu$ becomes brighter when moving from low-$M_{\rm BH}$ (blue profiles) to high-$M_{\rm BH}$ (red profiles). 

In our $M_{\rm BH}$--$\mu_{\rm 0, 3.6\mu m, sph}$ diagram (Figure \ref{Mbh_I0_Ie}), the core-S\'ersic galaxies are represented with the $\mu_{\rm 0, 3.6\mu m, sph}$ value that they presumably would originally have if their cores did not undergo a depletion\footnote{\label{def_core} The deficit of light at the center of core-S\'ersic galaxies is generally only a small fraction \citep[$5\%$, average value from table 5 in][]{Dullo:2019} of their total spheroid light.  This fraction is variable and can be approximately quantified for a given $M_{\rm *,sph}$ if we combine the  $M_{\rm BH}$--$M_{\rm *,sph}$ relation \citep[e.g., from][]{Sahu:2019:a} with the $M_{\rm BH}$--$M_{\rm *,def}$ relation from the literature \citep[e.g.,][]{Graham:2004, Ferrarese:2006, Dullo:2014, Savorgnan:Graham:2015}.}  of light due to coalescing BH binaries in dry major-mergers \citep{Begelman:1980}. 
Hence, one should not use the above relation to estimate $M_{\rm BH}$ using the actual (depleted) central surface brightness ($\mu_{\rm 0,core}$) for cored galaxies, instead, the $\mu_{\rm 0}$ extrapolated from the S\'ersic potion of their spheroid profile can be used. 
The actual $\mu_{\rm 0,core}$ for the core-S\'ersic galaxies dims with increasing $M_{\rm *,gal}$  \citep{Graham:Guzman:2003}. 


To include our remaining (non-Spitzer) sample of 22 galaxies, we mapped the central surface brightness, $\mu_{\rm 0, sph}$, values  to the central surface stellar mass density ($\Sigma_{\rm 0,sph}$) with the units of solar mass per square parsec  ($\rm M_\odot \, pc^{-2}$) using Equation \ref{Sigma_R}. 
We obtained a positive log-linear $M_{\rm BH}$--$\Sigma_{\rm 0,sph}$ relation, which is represented in the right-hand panel of Figure \ref{Mbh_I0_Ie}. 
The best-fit relation is provided in Table \ref{fit parameters1}, along with the (intrinsic\footnote{It should be noted that this depends on the adopted parameter uncertainties.} and total)  rms scatter, Pearson correlation coefficient, and Spearman rank-order correlation coefficient. The $M_{\rm BH}$--$\mu_{\rm 0, 3.6\mu m, sph}$ and $M_{\rm BH}$--$\Sigma_{\rm 0,sph}$ relations obtained using only S\'ersic galaxies are consistent with the relations obtained when including the core-S\'ersic galaxies. 

\begin{deluxetable*}{lclcccc}[t]
\tabletypesize{\footnotesize}
\tablecolumns{7}
\tablecaption{Correlations between the Black Hole Mass and the Spheroid Projected Density \label{fit parameters1}}
\tablehead{
\colhead{ \textbf{Category} } & \colhead{ \textbf{Number} } & \colhead{ \bm{$\log (M_{\rm BH}/M_\odot)=(Slope)\, X + (Intercept)$} } & \colhead{ \bm{$\epsilon$} }  & \colhead{ \bm{ $\Delta_{\rm  rms | BH}$} }  & \colhead{ \bm{$r_p$ } } & \colhead{ \bm{$r_s$}}   \\
\colhead{} & \colhead{} &  \colhead{\textbf{dex}} & \colhead{\textbf{dex}} & \colhead{\textbf{dex}} & \colhead{} & \colhead{}   \\
\colhead{ \textbf{(1)}} & \colhead{\textbf{(2)}} & \colhead{\textbf{(3)}} & \colhead{\textbf{(4)}} & \colhead{\textbf{(5)}} & \colhead{\textbf{(6)}} & \colhead{\textbf{(7)}} 
}
\startdata
& & \textbf{Central Surface Brightness (Figure \ref{Mbh_I0_Ie}, left-hand panel)}&  & & & \\
\hline
$\rm 3.6 \, \mu$m sample & 94\tablenotemark{a}  & $\log \left(M_{\rm BH}/\rm M_{\sun} \right ) = (-0.41 \pm 0.04)\, [\mu_{\rm 0, 3.6 \mu m, sph} - 13\, \rm mag \,arcsec^{-2}] + (7.97 \pm 0.10)$ & 1.00 & 1.03   & -0.52   & -0.51 \\
\hline
& & \textbf{Central Projected Mass Density (Figure \ref{Mbh_I0_Ie}, right-hand panel)}&  & & & \\
\hline
All types & 116\tablenotemark{a}  & $\log \left( M_{\rm BH}/\rm M_{\sun} \right )=(0.91 \pm 0.06) \, \log \left( \rm \Sigma_{\rm 0,sph}/10^6 \, M_{\sun} pc^{-2} \right)+(8.38 \pm 0.09)$ & 0.92  & 0.95 &  0.57    & 0.58 \\ 
\hline
& & \textbf{Projected density within 1 kpc (Figure \ref{Compactness}, left-hand panel)} &  & & & \\
\hline
\vspace*{1mm}
All types & 119  & $\log \left( M_{\rm BH}/\rm M_{\sun}  \right )=(2.69 \pm 0.18) \, \log \left( \rm \langle \Sigma \rangle_{\rm 1 kpc,sph}/10^{3.5} \, M_{\sun} pc^{-2} \right)+(7.84 \pm 0.07)$  & 0.57 & 0.69 &  0.78   & 0.80  \\ 
\hline
& & \textbf{Projected density within 5 kpc (Figure \ref{Compactness}, right-hand panel)} &  & & & \\
\hline
\vspace*{1mm}
All types & 119  & $\log \left( M_{\rm BH}/\rm M_{\sun}  \right )=(1.87 \pm 0.10) \, \log \left( \rm \langle \Sigma \rangle_{\rm 5 kpc,sph}/10^{2} \, M_{\sun} pc^{2.5} \right)+(8.13 \pm 0.05)$ & 0.51 & 0.59 &  0.83   & 0.84\\ 
\hline
& &  \textbf{Effective Surface Brightness at $R_{ \rm e, sph}$ (Figure \ref{Mbh_Ie}, left-hand panel)} &  & & & \\
\hline
\vspace*{1mm}
LTGs ($\rm 3.6 \, \mu$m sample) & 26  & $\log \left( M_{\rm BH}/\rm M_{\sun} \right )=(0.77 \pm 0.12)\, [\mu_{\rm e, 3.6 \mu m, sph} - 19\, \rm mag \,arcsec^{-2}] + (7.84 \pm 0.17)$ & 0.75 & 0.85 &  0.51   & 0.54 \\ 
\vspace*{1mm}
ETGs ($\rm 3.6 \, \mu$m sample) & 71  & $\log \left( M_{\rm BH}/\rm M_{\sun}  \right )=(0.47 \pm 0.04)\, [\mu_{\rm e, 3.6 \mu m, sph} - 19\, \rm mag \,arcsec^{-2}] + (7.95 \pm 0.11)$ & 0.78 & 0.83 &  0.57   & 0.57 \\ 
\vspace*{1mm}
E ($\rm 3.6 \, \mu$m sample) & 35  & $\log \left( M_{\rm BH}/\rm M_{\sun}  \right )=(0.97 \pm 0.10)\, [\mu_{\rm e, 3.6 \mu m, sph} - 19\, \rm mag \,arcsec^{-2}] + (6.24 \pm 0.34)$ & 1.03 & 1.14 &  0.05   & 0.08 \\ 
\vspace*{1mm}
ES/S0 ($\rm 3.6 \, \mu$m sample) & 36  & $\log \left( M_{\rm BH}/\rm M_{\sun}  \right )=(0.76 \pm 0.10)\, [\mu_{\rm e, 3.6 \mu m, sph} - 19\, \rm mag \,arcsec^{-2}] + (8.41 \pm 0.15)$ & 0.81 & 0.92 &  0.36   & 0.35 \\ 
\hline
& & \textbf{Projected Density at $ R_{\rm e, sph}$ (Figure \ref{Mbh_Ie}, middle panel)}  &  & & & \\
\hline
\vspace*{1mm}
LTGs & 39  & $\log \left( M_{\rm BH}/\rm M_{\sun} \right )=(-1.56 \pm 0.22) \, \log \left( \rm \Sigma_{\rm e,sph}/10^3 \, M_{\sun} pc^{-2} \right)+(7.75 \pm 0.15)$ & 0.69 & 0.77 &  -0.43   & -0.43 \\ 
\vspace*{1mm}
ETGs & 80  & $\log \left( M_{\rm BH}/\rm M_{\sun}  \right )=(-1.11 \pm 0.08) \, \log \left( \rm \Sigma_{\rm e,sph}/10^3 \, M_{\sun} pc^{-2} \right)+(8.31 \pm 0.09)$ & 0.76 & 0.81 &  -0.55   & -0.52 \\ 
\vspace*{1mm}
E & 40  & $\log \left( M_{\rm BH}/\rm M_{\sun}  \right )=(-1.01 \pm 0.39) \, \log \left( \rm \Sigma_{\rm e,sph}/10^3 \, M_{\sun} pc^{-2} \right)+(8.14 \pm 0.35)$ & 0.73 & 0.76 &  -0.004   & -0.01 \\ 
\vspace*{1mm}
ES/S0 & 40  & $\log \left( M_{\rm BH}/\rm M_{\sun}  \right )=(-1.60 \pm 0.27) \, \log \left( \rm \Sigma_{\rm e,sph}/10^3 \, M_{\sun} pc^{-2} \right)+(8.78 \pm 0.18)$ & 0.76 & 0.85 &  -0.35   & -0.32 \\ 
\hline
& & \textbf{Projected Density within $R_{\rm e, sph}$ (Figure \ref{Mbh_Ie}, right-hand panel)} &  & & & \\
\hline
\vspace*{1mm}
LTGs & 39  & $\log \left( M_{\rm BH}/\rm M_{\sun}  \right )=(-1.69 \pm 0.28) \, \log \left( \rm \langle \Sigma \rangle_{\rm e,sph} /10^3 \, M_{\sun} pc^{-2} \right)+(8.45 \pm 0.27)$ & 0.71 & 0.80 &  -0.37  & -0.40 \\ 
\vspace*{1mm}
ETGs & 80  & $\log \left( M_{\rm BH}/\rm M_{\sun}  \right )=(-1.24 \pm 0.10) \, \log \left( \rm \langle \Sigma \rangle_{\rm e,sph}/10^3 \, M_{\sun} pc^{-2} \right)+(8.96 \pm 0.10)$ & 0.76 & 0.82 &  -0.53   & -0.50 \\ 
\vspace*{1mm}
E & 40  & $\log \left( M_{\rm BH}/\rm M_{\sun}  \right )=(-1.06 \pm 0.59) \, \log \left( \rm \langle \Sigma \rangle_{\rm e,sph}/10^3 \, M_{\sun} pc^{-2} \right)+(8.75 \pm 0.18)$ & 0.71 & 0.74 &  -0.02   & -0.02 \\ 
\vspace*{1mm}
ES/S0 & 40  & $\log \left( M_{\rm BH}/\rm M_{\sun}  \right )=(-1.62 \pm 0.38) \, \log \left( \rm \langle \Sigma \rangle_{\rm e,sph}/10^3 \, M_{\sun} pc^{-2} \right)+(9.54 \pm 0.37)$ & 0.78 & 0.87 &  -0.27   & -0.25 \\ 
\enddata
\tablecomments{
Columns:
(1) Galaxy type.
(2) Number of galaxies.
(3) Scaling relation obtained from the \textsc{bces(bisector)} regression.
(4) Intrinsic scatter in the $\log M_{\rm BH}$-direction \citep[using Equation 1 from][]{Graham:Driver:2007}.
(5) Total root mean square (rms) scatter in the $\log M_{\rm BH}$ direction.
(6) Pearson correlation coefficient. 
(7) Spearman rank-order correlation coefficient. 
}
\tablenotetext{a}{Regression performed after excluding three outliers, see Section \ref{3.1} for more details.}
\end{deluxetable*}

\subsection{Projected Mass Density within 1 kpc: The Spheroid Compactness $\langle \Sigma \rangle_{\rm 1 kpc,sph}$}
\label{3.2}

\begin{figure*}
\begin{center}
\includegraphics[clip=true,trim= 10mm 01mm 18mm 12mm,width=   0.99\textwidth]{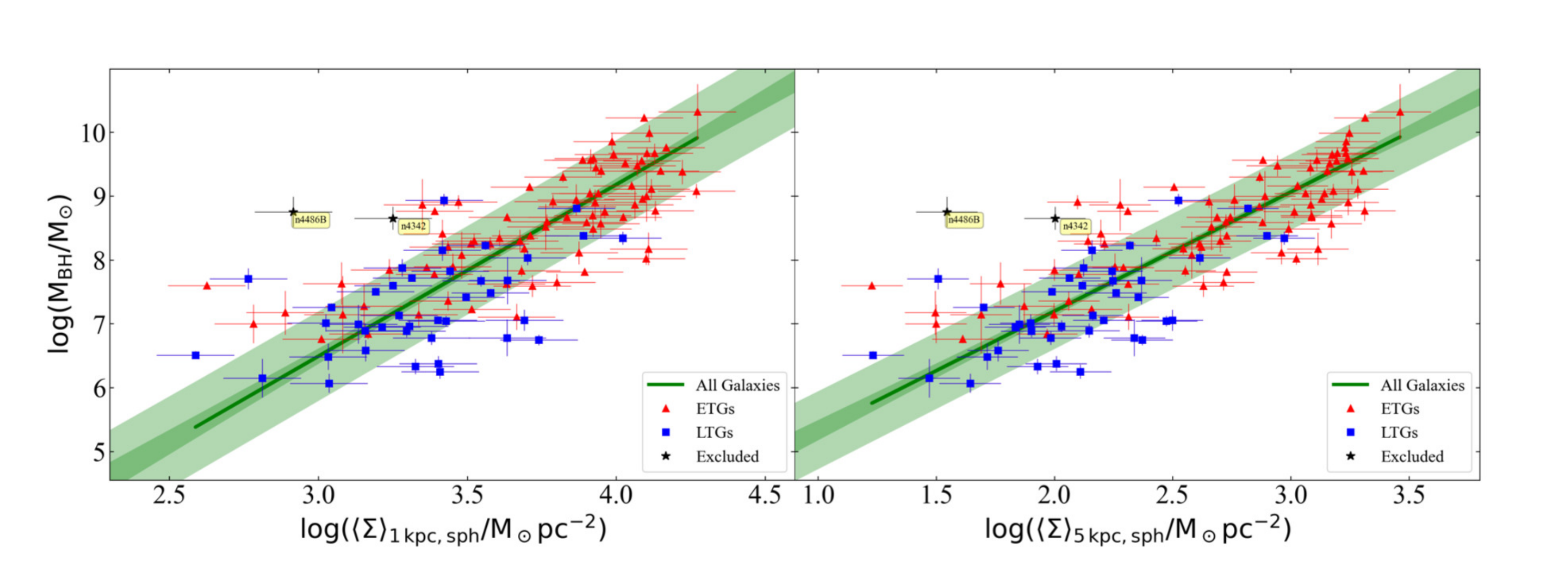} 
\caption{Black hole mass plotted against the spheroid's projected stellar mass density within the inner 1 kpc (left-hand panel, Equation \ref{Mbh_Sigma_1kpc}) and the spheroid's projected density within 5 kpc (right-hand panel, Equation \ref{Mbh_Sigma_5kpc}). ETGs and LTGs are marked differently to depict that the two types follow the same relation.}
\label{Compactness}
\end{center}
\end{figure*}

\begin{figure}
\begin{center}
\includegraphics[clip=true,trim= 11mm 05mm 18mm 12mm,width=   0.5\textwidth]{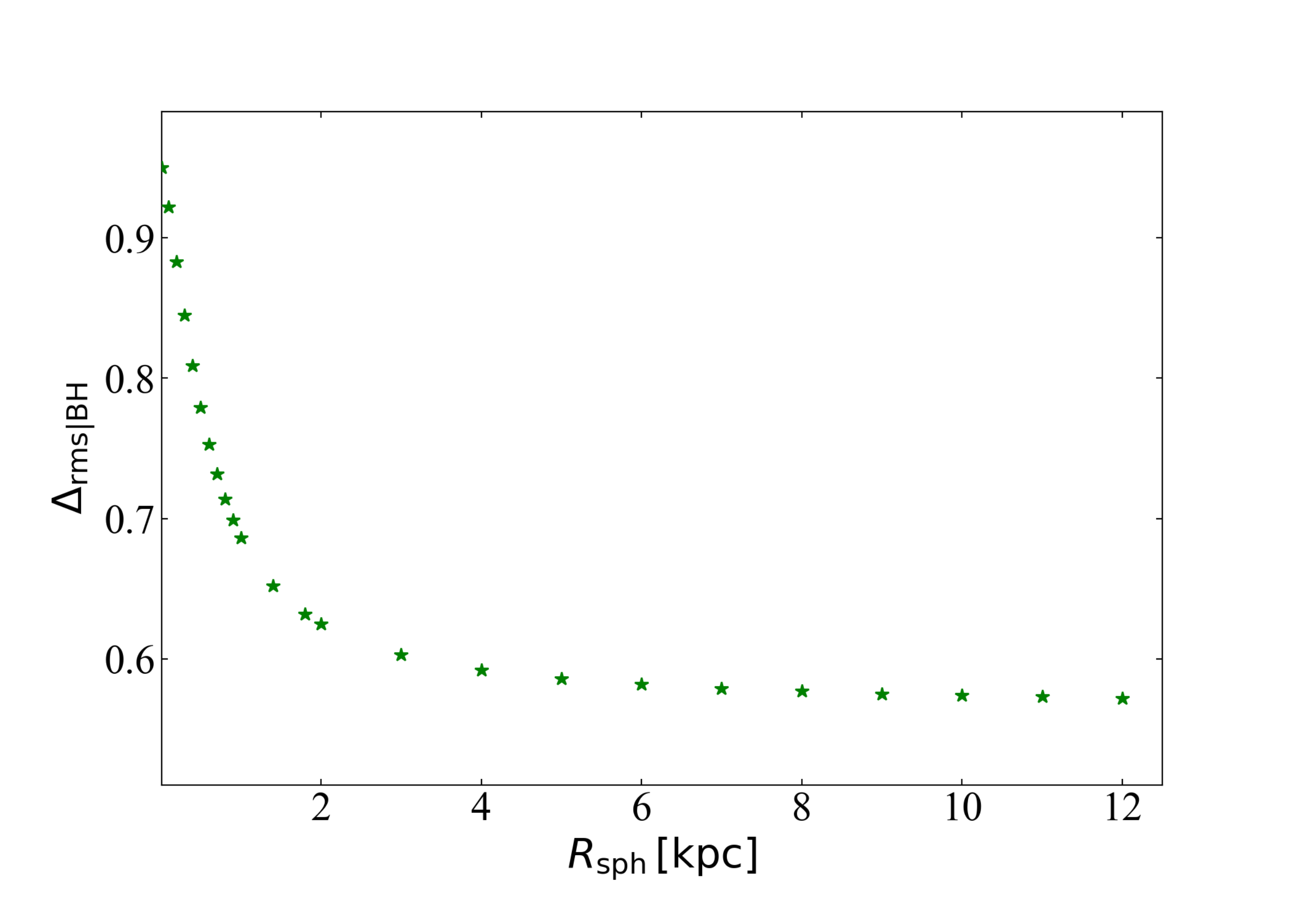} 
\caption{The total vertical rms scatter in the $M_{\rm BH}$--$\langle \Sigma \rangle_{\rm R,sph}$ relations as a function of $R_{\rm sph}$.}
\label{Scatter_R_Sigma}
\end{center}
\end{figure}

The projected stellar mass density ($\langle \Sigma \rangle_{\rm 1 kpc}$) within the inner 1 kpc of a galaxy has been used as a measure of galaxy \enquote{compactness} \citep{Barro:Faber:2017, Ni:Brandt:2020}, and  to identify compact star forming galaxies  \citep{Suess:Kriek:2021}. 
Interestingly, for star-forming galaxies $\langle \Sigma \rangle_{\rm 1 kpc}$ has been found to correlate with the black hole growth\footnote{The connection between black hole growth and the host galaxy's stellar mass density is explained by the assumption of a linear correlation between stellar density and gas density for star-forming galaxies \citep[see][ and references therein]{Lin:Pan:2019}.  Thus, a high value of $\langle \Sigma \rangle_{\rm 1 kpc}$ for a star-forming galaxy infers a high gas density, and the abundance of gas at the inner galactic regions is known to boost the black hole growth \citep{Dekel:Lapiner:2019, Habouzit:Genel:2019}.}, and it has been suggested that this correlation is stronger than the connection between the black hole growth and host galaxy stellar mass \citep[see][]{Ni:Brandt:2020}. 
Additionally, it has been suggested that $\langle \Sigma \rangle_{\rm 1 kpc}$ is a better indicator of black hole growth than the projected stellar mass density within the galaxy half-light radius and the projected density within other (smaller or larger) constant (e.g., 0.1 kpc, 10 kpc)  radii \citep[][]{Ni:Yang:2019, Ni:Brandt:2020}.


Here we investigate a possible correlation between the black hole mass and the average projected stellar mass density  ($\langle \Sigma \rangle_{\rm 1 kpc,sph}$) within the inner 1 kpc of the host spheroid.
Thus, we refer to $\langle \Sigma \rangle_{\rm 1 kpc,sph}$ as the spheroid compactness.
Most of our sample with directly-measured  $M_{\rm BH}$ are quiescent, with $\langle \Sigma \rangle_{\rm 1 kpc,sph}$ greater than the  critical/threshold value \citep[$\rm \langle \Sigma \rangle_{\rm 1 kpc}=3 \times 10^3 \, M_\odot \, pc^{-2}$,][]{Cheung:Faber:2012} used to identify the quiescent galaxies \citep{Hopkins:Wellons:2021}. 
Additionally, we explored how the correlation between $M_{\rm BH}$ and spheroid compactness compares against the correlation between $M_{\rm BH}$ and the spheroid densities at/within other radii.

We find a tight $M_{\rm BH}$--$\langle \Sigma \rangle_{\rm 1 kpc,sph}$ correlation (left-hand panel in Figure \ref{Compactness}), where all galaxy types (ETGs+LTGs) seem to follow a single positive relation\footnote{Here,  we use a $30\%$ (0.13 dex) uncertainty on the $\langle \Sigma \rangle_{\rm 1 kpc,sph}$ (and the later discussed $\langle \Sigma \rangle_{\rm 5 kpc,sph}$) values.  Consistent relations are obtained when using up to a $40\%$ (0.17 dex) uncertainty.}, such that 
\begin{IEEEeqnarray}{rCl}
\label{Mbh_Sigma_1kpc}
\rm  \log \left( \frac{M_{\rm BH}}{\rm M_\odot } \right) &=& \rm (2.69 \pm 0.18) \log \left( \frac{\langle \Sigma \rangle_{\rm 1 kpc,sph}}{\rm 10^{3.5} \, M_\odot \, pc^{-2}} \right)  \nonumber \\
&& +\> (7.84 \pm 0.07), 
\end{IEEEeqnarray}
with $\Delta_{\rm rms|BH}=0.69$ dex (see Table \ref{fit parameters1} for correlation coefficients). 
Similarly, we also see a positive trend between $M_{\rm BH}$ and the column (stellar mass) density within other projected spheroid radii (e.g., 0.01 kpc, 0.1 kpc, 5 kpc, 10 kpc). This positive trend is evident from the distribution of spheroid profiles in the left-hand panel of Figure \ref{Sersic_all}, where, at all fixed radii, the redder profiles with higher $M_{\rm BH}$ are brighter than the bluer profiles with lower $M_{\rm BH}$.

The correlation between $M_{\rm BH}$ and densities within fixed physical radii smaller than 1 kpc ($\langle \Sigma \rangle_{\rm 0.01 kpc,sph}$  and $\langle \Sigma \rangle_{\rm 0.1 kpc,sph}$) are not as tight as the above relation (Equation \ref{Mbh_Sigma_1kpc}). However, we find better $M_{\rm BH}$--$\langle \Sigma \rangle_{\rm R,sph}$ correlations for $\rm R > 1\, kpc$,  with a gradually shallower slope and smaller scatter than the $M_{\rm BH}$--$\langle \Sigma \rangle_{\rm 1 kpc,sph}$ relation. 
For  example, the relation  between $M_{\rm BH}$ and $\langle \Sigma \rangle_{\rm 5 kpc,sph}$ shown in the right-hand panel of Figure \ref{Compactness}. It can be expressed as,  
\begin{IEEEeqnarray}{rCl}
\label{Mbh_Sigma_5kpc}
\rm  \log \left( \frac{M_{\rm BH}}{\rm M_\odot } \right) &=& \rm (1.87 \pm 0.10) \log \left( \frac{\langle \Sigma \rangle_{\rm 5 kpc,sph}}{\rm 10^{2.5} \, M_\odot \, pc^{-2}} \right)  \nonumber \\
&& +\> (8.13\pm 0.05), 
\end{IEEEeqnarray}
with the total rms scatter $\Delta_{\rm rms|BH}=0.59$ dex. A plot of the rms scatter  about the $M_{\rm BH}$--$\langle \Sigma \rangle_{\rm R,sph}$ relation as a function of  R is shown in Figure \ref{Scatter_R_Sigma}. The scatter asymptotes to  $\sim 0.58\pm 0.01$ dex beyond 5 kpc.

The low (0.69 dex) scatter about the $M_{\rm BH}$--$\langle \Sigma \rangle_{\rm 1 kpc,sph}$ relation relative to the 0.95 dex scatter about the  $M_{\rm BH}$--$\Sigma_{\rm 0,sph}$ relation (Table \ref{fit parameters1}) and the (soon to be discussed) $M_{\rm BH}$--$\langle \Sigma \rangle_{\rm e,sph}$ relations suggests that $\langle \Sigma \rangle_{\rm 1kpc,sph}$ is a  better predictor of $M_{\rm BH}$ than the latter projected  mass densities. However, the $M_{\rm BH}$--$\langle \Sigma \rangle_{\rm R,sph}$ relations for  $\rm R> 1$ kpc is stronger,  reflective of the separation of the  $\langle \mu \rangle$ (and $\langle \Sigma \rangle$) profiles at large radii.

\subsection{Surface Brightness and Projected Density at the Half-Light Radius: $\mu_{\rm e, 3.6 \mu m, sph} \, \& \, \Sigma_{\rm e,sph}$}
\label{3.3}
\begin{figure*}
\begin{center}
\includegraphics[clip=true,trim= 06mm 01mm 13mm 11mm,width= 1\textwidth]{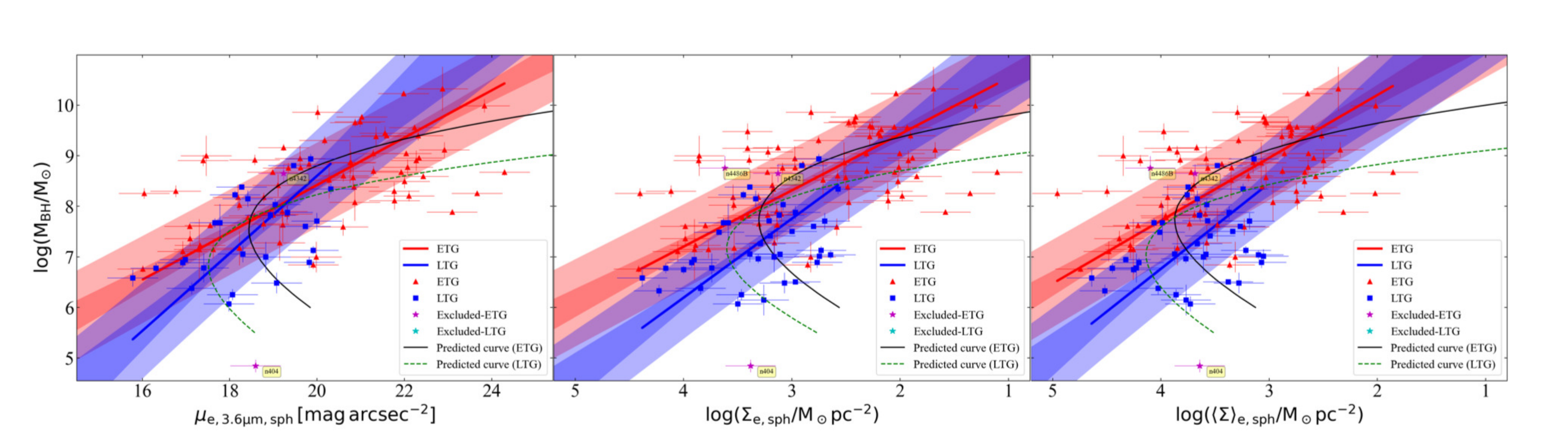}
\caption{Black hole mass versus the bulge surface brightness $\mu_{\rm e, 3.6 \mu m, sph}$ at $R_{\rm e,sph}$ (left-hand panel, Equations \ref{Mbh_mue_ETGs} and \ref{Mbh_mue_LTGs}), the projected stellar mass density $\Sigma_{\rm e,sph}$ at $R_{\rm e,sph}$ (middle panel, Table \ref{fit parameters1}), and the average stellar mass density $\langle \Sigma \rangle_{\rm e,sph}$ within $R_{\rm e,sph}$ (right-hand panel, Table \ref{fit parameters1}). 
ETGs and LTGs seem to define different relations in these diagrams. 
However, the complete picture of these relations is curved, as shown by the expected black and green curves for ETGs and LTGs, respectively. 
Note that the horizontal-axes in the middle and right-hand panels are inverted, and in the first panel, the horizontal-axis presents $\mu_{\rm e, 3.6 \mu m, sph}$ in a dimming order from left to right.
}
\label{Mbh_Ie}
\end{center}
\end{figure*} 

Using our $\rm 3.6\, \mu m$-sample, we see a positive trend between $M_{\rm BH}$ and the surface brightness ($\mu_{\rm e, sph}$) at the projected half-light radius of spheroids (Figure \ref{Mbh_Ie}).  
A higher magnitude of $\mu_{\rm e, sph}$ corresponds to a lower luminosity density; thus,  we find a declining relation between $M_{\rm BH}$ and the effective luminosity density.
 We observe that ETGs and LTGs in our sample define two different $M_{\rm BH}$--$\mu_{\rm e, 3.6 \mu m, sph}$ relations, which are represented in the left-hand panel of Figure \ref{Mbh_Ie}. 
ETGs define the following relation,
\begin{IEEEeqnarray}{rCl}
\label{Mbh_mue_ETGs}
\rm \log \left( \frac{M_{\rm BH}}{\rm M_\odot } \right) &=& \rm (0.47 \pm 0.04)\, \left[\mu_{\rm e, 3.6\mu m, sph} - \frac{19\, mag}{arcsec^{2}} \right] \nonumber \\
&& \> +\, (7.95 \pm 0.11), 
\end{IEEEeqnarray}
 with $\Delta_{\rm rms|BH}=0.83$ dex. Whereas the LTGs follow a steeper relation given by
\begin{IEEEeqnarray}{rCl}
\label{Mbh_mue_LTGs}
\rm \log \left( \frac{M_{\rm BH}}{\rm M_\odot } \right) &=& \rm (0.77 \pm 0.12)\, \left[\mu_{\rm e, 3.6\mu m, sph} - \frac{19\, mag}{arcsec^{2}} \right]  \nonumber \\
&& \> +\, (7.84 \pm 0.17), 
\end{IEEEeqnarray}
with $\Delta_{\rm rms|BH}=0.85$ dex. 

In order to include our full sample, we mapped $\mu_{\rm e, sph}$ ($\rm mag \, arcsec^{-2}$) to $\Sigma_{\rm e,sph}$ ($\rm M_\odot pc^{-2}$) using Equation \ref{Sigma_R}, and recover two  trends defined by ETGs and LTGs in the  $M_{\rm BH}$--$\Sigma_{\rm e,sph}$ diagram. 
Similar trends due to ETGs and LTGs are observed in the $M_{\rm BH}$--($\langle \Sigma \rangle_{\rm e,sph}$, average projected density within $R_{\rm e,sph}$) diagram. 
The $M_{\rm BH}$--$\Sigma_{\rm e,sph}$ and $M_{\rm BH}$--$\langle \Sigma \rangle_{\rm e,sph}$ relations are depicted, respectively, in the middle and the right-hand panel of Figure \ref{Mbh_Ie}. 
The fit parameters and the correlation coefficients for these distributions are provided in Table \ref{fit parameters1}. 

For early-type galaxies, the galaxy luminosity (or mass)  has a curved relation with the galaxy surface brightness at/within any scale radius, $R_{\rm z, sph}$, enclosing (a non-zero) $\rm z\%$ of the galaxy's total light \citep{Graham:Re:2019}. 
This includes the relation between galaxy luminosity and galaxy surface brightness at/within the half-light  ($\rm z=50\%$) radius, i.e., $M_{\rm *, gal}$--$\mu_{\rm e, gal}$ or $M_{\rm *, gal}$--$\langle \mu \rangle_{\rm e, gal}$ \citep[see][their figure 3]{Graham:Re:2019}.
Similarly for spheroids, the $M_{\rm *, sph}$--$\mu_{\rm e, sph}$ and  $M_{\rm BH}$--$\mu_{\rm e, sph}$ (also $\Sigma_{\rm e,sph}$, and $\langle \Sigma \rangle_{\rm e,sph}$) relations are expected to be curved, as shown in Figure \ref{Mbh_Ie}. 
These curved relations for ETGs and LTGs  are predicted using the  $M_{\rm BH}$--$n_{\rm sph}$ \citep{Sahu:2019:a}  and $M_{\rm BH}$--$\mu_{\rm 0, 3.6 \mu m, sph}$ relations defined by the two morphological classes for the $M_{\rm BH}$ range of our sample, and further applying the equation $\mu_{\rm e, sph}=\mu_{\rm 0, sph}+ \rm 2.5 \, b_n/ln(10)$ (or $\Sigma_{\rm e,sph}=\Sigma_{\rm 0,sph}- \rm b_n/ln(10)$) for a S\'ersic distribution. 
Here, $\langle \Sigma \rangle_{\rm e,sph}$, $\langle \mu \rangle_{\rm e,sph}$, and $\mu_{\rm e, sph}$ can be related through equation 9 in \citep{Graham:Driver:2005} and equation 11 in \citep{Graham:Merritt:2006}.

Thus, the slopes of the fitted  $M_{\rm BH}$--$\mu_{\rm e, sph}$  lines (e.g., Equations \ref{Mbh_mue_ETGs} and \ref{Mbh_mue_LTGs}) obtained here depend on the  $M_{\rm BH}$ and $\mu_{\rm e, sph}$  range of the fitted sample.
Moreover, as the ETGs and LTGs seem to follow different $M_{\rm BH}$--$\mu_{\rm e, 3.6 \mu m, sph}$  trends, the foreseeable $M_{\rm BH}$--$\mu_{\rm e, 3.6 \mu m, sph}$ (or $\Sigma_{\rm e,sph}$, or $\langle \Sigma \rangle_{\rm e,sph}$) curves shall also be different for the two galaxy types. 


\begin{figure*}
\begin{center}
\includegraphics[clip=true,trim= 06mm 01mm 13mm 11mm,width= 1\textwidth]{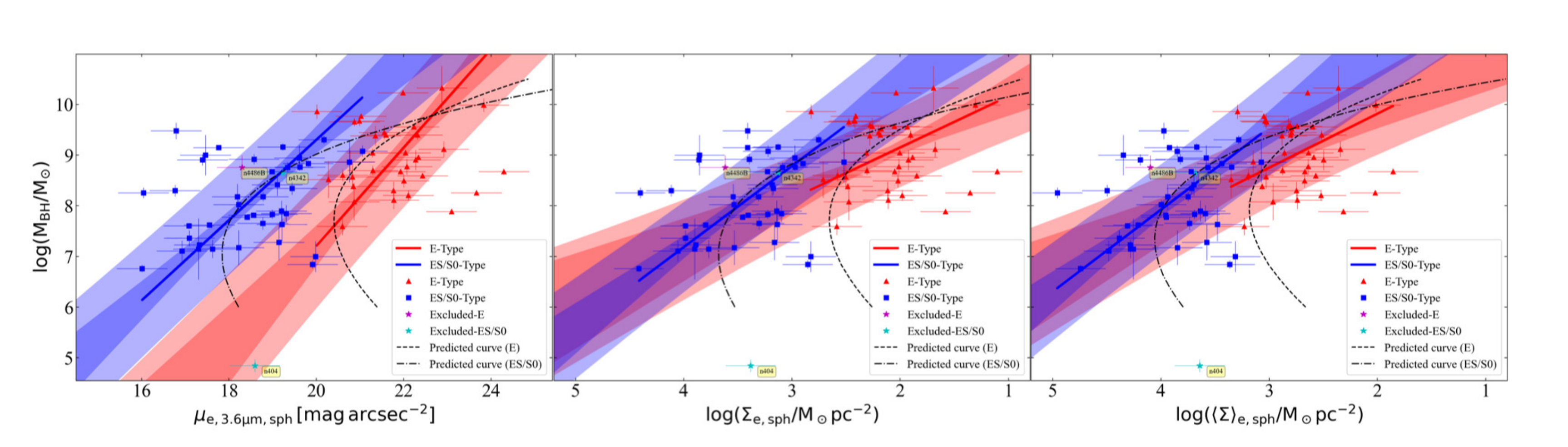}
\caption{Similar to Figure \ref{Mbh_Ie}, but now showing different regressions performed for   ETGs with a disk (ES- and S0-types) and ETGs without a disk (E-type). 
The correlation parameters are provided in Table \ref{fit parameters1}. The dashed and dot-dashed curves represent the expected relations for  E-type and ES/S0-type galaxies, respectively. 
}
\label{Mbh_Ie_E_ESS0}
\end{center}
\end{figure*} 
 
\subsubsection{Offset between ETGs with and without a disk}

\citet{Sahu:2019:a} observed an offset  of $1.12\pm 0.20$ dex in the $M_{\rm BH}$-direction, between ETGs with a disk (ES- and S0-types) and ETGs without a disk (E-type), which defined almost parallel relations in the $M_{\rm BH}$--$M_{\rm *, sph}$ diagram. 
The calculation of $M_{\rm *, sph}$  is based on the spheroid S\'ersic profile, quantified by the parameters  $n_{\rm sph}$, $R_{\rm e, sph}$, and $\mu_{\rm e, 3.6 \mu m, sph}$, and thus the offset between ES/S0- and E-types must have propagated to $M_{\rm *, sph}$  from these parameters. 
Further, \citet{Sahu:2020} re-observed this offset ($1.38\pm 0.28$ dex in the $M_{\rm BH}$-direction) between ETGs with and without a disk in the $M_{\rm BH}$--$R_{\rm e, sph}$ diagram, where again these categories defined almost parallel relations.  \citep{Sahu:2020} did not report any significant offset between ETG subsamples in the $M_{\rm BH}$--$n_{\rm sph}$ diagram. 
Here, we next investigated if there is any such offset between ES/S0- and E-types in the $M_{\rm BH}$--$\mu_{\rm e, 3.6 \mu m, sph}$ diagram. 

Upon separating the ETGs with and  without a disk, we do see the two groups offset from each other in the $M_{\rm BH}$--$\mu_{\rm e, 3.6 \mu m, sph}$,  $M_{\rm BH}$--$\Sigma_{\rm e,sph}$, and $M_{\rm BH}$--$\langle \Sigma \rangle_{\rm e,sph}$ diagrams (Figure \ref{Mbh_Ie_E_ESS0}). 
However, the quality of fit for the two samples is poor (see Table \ref{fit parameters1}); thus, it is difficult to quantify the offset accurately. 
Moreover, as discussed before, the complete $M_{\rm BH}$--$\mu_{\rm e, 3.6 \mu m, sph}$ relations are curved. 
The expected $M_{\rm BH}$--$\mu_{\rm e, 3.6 \mu m, sph}$  curves for E- and  ES/S0-types are shown in Figure \ref{Mbh_Ie_E_ESS0}. These curves are also calculated  by using the $M_{\rm BH}$--$n_{\rm sph}$ and $M_{\rm BH}$--$\mu_{\rm 0, sph}$ lines for the two populations combined with  $\mu_{\rm e, sph}=\mu_{\rm 0, sph}+ \rm 2.5 \, b_n/ln(10)$.
Additionally, as these curves are not parallel, the offset between the relations for E-type and ES/S0-type may not be constant throughout. 

The declining $M_{\rm BH}$--(effective surface brightness) relation can also be inferred from the distribution of the spheroid surface brightness profiles for our sample shown in the right-hand panel of Figure \ref{Sersic_all}. 
At the half-light radii ($\rm R/R_e = 1$) of spheroids, the  surface brightness dims when going from low-$M_{\rm BH}$   (blue) to high-$M_{\rm BH}$  (red) profiles. 
Also, given the radially declining projected density profile, the ES/S0-types with a smaller $R_{\rm e, sph}$ than the E-types, have a brighter $\mu_{\rm e,sph}$ (higher $\Sigma_{\rm e,sph}$ and $\langle \Sigma \rangle_{\rm e,sph}$) than E-types hosting similar $M_{\rm BH}$.  
This is why the direction of the offset between E- and ES/S0-types in these diagrams (Figure \ref{Mbh_Ie_E_ESS0}) is  opposite to  the offset seen in the $M_{\rm BH}$--$R_{\rm e, sph}$ and $M_{\rm BH}$--$M_{\rm *, sph}$ diagrams, where ES/S0-types have a smaller $R_{\rm e, sph}$ and $M_{\rm *, sph}$ than the E-types hosting a similar $M_{\rm BH}$.

\section{Black Hole Mass versus Spheroid Spatial Density}
\label{Deprojected_density}

\begin{figure*}
\begin{center}
\hspace*{-0.8cm}
\includegraphics[clip=true,trim= 05mm 05mm 010mm 08mm,width=   1.1\textwidth]{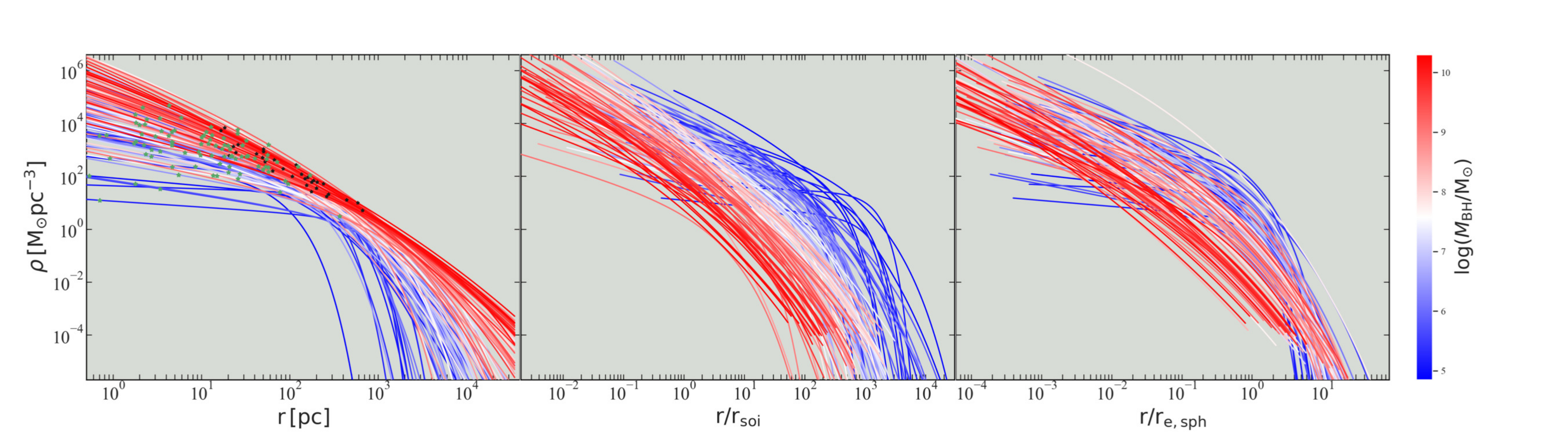}
\caption{Spheroid internal stellar mass density profiles. 
The sequential color map blue-white-red depicts an increasing order of $M_{\rm BH}$. 
In the left-hand panel, the black and green colored stars mark the black hole's influence radius for the core-S\'ersic and S\'ersic galaxies, respectively.
The horizontal axes are scaled with respect to the sphere of influence radius ($r_{soi}$) of black holes and the spheroid's spatial half-light radius $r_{e}$ for the profiles shown in the middle and right-hand panels, respectively.
}
\label{Rho_profiles}
\end{center}
\end{figure*}

We deprojected the (equivalent-axis) S\'ersic surface brightness profiles of the spheroids to obtain their spatial (i.e., internal) mass density profiles, as described in the Appendix  \ref{calculation_density}. 
These internal density profiles\footnote{For our black hole correlations, the internal densities are numerically calculated using the exact integral expressed by Equation \ref{Rho_int}. However, the extended internal density profiles in Figure \ref{Rho_profiles} are calculated using an approximated model \citep{Prugniel1997}. This is because, for some spheroids, the density integral (Equation \ref{Rho_int})  did not converge to provide a valid/real density value, especially at larger radii. Moreover, using the approximate model can still explain the qualitative nature of the $M_{\rm BH}$--$\rho$ trends observed here.}  are displayed in Figure \ref{Rho_profiles}. 
We used a sequential blue-white-red color map to represent the central black hole masses in increasing order from low-mass (blue) to high-mass (red).  
The density profiles in the three panels of Figure \ref{Rho_profiles} will help one understand the upcoming correlations observed between the black hole mass and the host spheroid's internal density at various radii. 

Similar to the projected surface brightness profiles, the (deprojected) internal density profiles, $\rho(r)$, are monotonically declining and can be characterized using the S\'ersic surface brightness profile parameters ($n$, $R_{\rm e}$, $\rm \mu_e=-2.5 \log I_e$, see Equation \ref{Rho_int}). Smaller and less massive spheroids, generally quantified by smaller S\'ersic parameters ($n$ and $R_{\rm e}$), have a shallow inner density profile that descends quickly at outer radii (see the bluer profiles in the left-hand panel of Figure \ref{Rho_profiles}). 
On the contrary,  more massive spheroids, generally indicated by higher S\'ersic parameters ($n$ and $R_{\rm e}$), have a steeper  inner density profile with a higher density and a shallower decline at large radii (see the red profiles in the left-hand panel of Figure \ref{Rho_profiles}).

The horizontal-axes in the middle and the right-hand panels of Figure \ref{Rho_profiles} are scaled using the sphere-of-influence radius ($r_{\rm soi}$) of the black holes and the internal (or spatial) half-mass radius ($ r_{\rm e, sph}$) of the spheroids, respectively. This accounts for some of the different size scales used and will help with the understanding of the observed ($M_{\rm BH}$)--(spheroid internal density) relations revealed in the following sub-sections.

\subsection{Spatial Density at the Black Hole's Sphere-of-Influence: $\rho_{\rm soi, sph}$}
\label{4.1}

  
Based on the exact deprojection of the S\'ersic model  (Equations \ref{Rho_int}) the internal  density near the spheroid center, $\rho(r \rightarrow 0)$, tends to infinity\footnote{For $n=1$, $\rho(r \rightarrow 0)$ tends to a finite value and tends to zero for $\rm n<1$ (see Equation \ref{Rho_int}). Whereas, based on the \citet{Prugniel1997} model (Equation \ref{PS}) which is an approximation of the exact deprojection,  $\rho(r \rightarrow 0)$ tends to infinity for $n \gtrsim 0.6$.} for $n>1$.
Hence, as a measure of the central internal density, we chose the internal density at another central radius, where the gravitational potential of the black hole is in dynamical equilibrium with that of the host galaxy, known as the sphere-of-influence radius ($r_{\rm soi}$) of the black hole. 
We denote the spheroid spatial density at $r_{\rm soi}$ by $\rho_{\rm soi, sph}$.

\begin{figure}
\begin{center}
\includegraphics[clip=true,trim= 05mm 02mm 1.5mm 10mm,width=   0.5\textwidth]{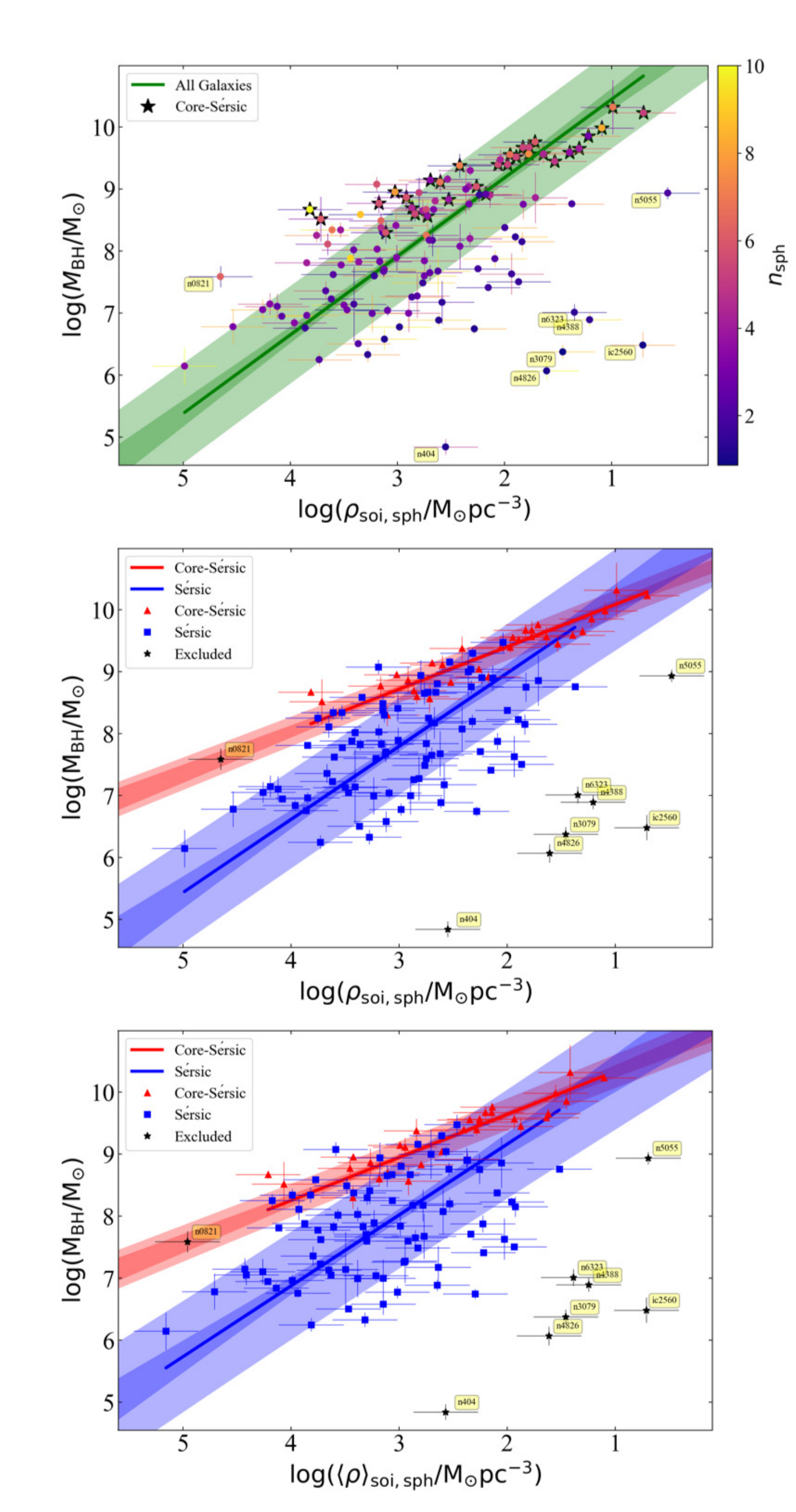}
\caption{Black hole mass versus internal stellar mass density at $r_{\rm soi}$ (top and middle panels) and black hole mass versus (averaged) internal stellar mass density within $r_{\rm soi}$ (bottom panel). 
The top panel shows a single regression (Footnote \ref{single_Mbh_Rho_soi}), where core-S\'ersic galaxies (black stars) are distributed in a manner that suggests a different $M_{\rm BH}$--$\rho_{\rm soi, sph}$ trend for these high-$n_{\rm sph}$ systems. 
All the data points in this panel are color-coded according to their S\'ersic indices.
The middle panel shows the two different $M_{\rm BH}$--$\rho_{\rm soi, sph}$  relations defined by S\'ersic (blue) and core-S\'ersic galaxies (red) galaxies. 
Similar substructure due to S\'ersic (low $n_{\rm sph}$) and core-S\'ersic (high $n_{\rm sph}$) galaxies are observed in the $M_{\rm BH}$--$\langle \rho \rangle_{\rm soi, sph}$ diagram (bottom panel).  The excluded galaxies are named. See the text in Section \ref{4.1} for details. Note that the horizontal axis is inverted such that the density decreases when going from left to right.
}
\label{Mbh-Rho_soi}
\end{center}
\end{figure}

We first calculated $r_{\rm soi}$ using the following standard definition \citep{Peebles:1972, Frank:Rees:1976, Merritt2004, Ferrarese:Ford:2005}, 
\begin{IEEEeqnarray}{rCl}
\label{r_soi}
\rm r_{soi}=  \frac{G\,M_{BH}}{\sigma^2}, 
\end{IEEEeqnarray}
where $\sigma$ is the host galaxy's central (projected) stellar velocity dispersion, which is likely to be dominated by the spheroid component of our galaxies. 
The stellar velocity dispersions of our galaxies are primarily taken from the \textsc{HyperLeda} \citep{Makarov2014}  database\footnote{The  stellar velocity dispersions available at the \textsc{HyperLeda} database are homogenized to a constant aperture size of $\sim 0.595$ kpc.}, and are listed in \citet[][their table 1]{Sahu:2019:b}.  
The value of  $\rho_{\rm soi, sph}$ was numerically calculated using the  Equation \ref{Rho_int} at $r=r_{\rm soi}$. We also include the core-S\'ersic galaxies in the $M_{\rm BH}$--$\rho_{\rm soi, sph}$ diagram (Figure \ref{Mbh-Rho_soi}), for whom $\rho_{\rm soi, sph}$ is based on the de-projection of the (inwardly extrapolated) S\'ersic component of their core-S\'ersic surface brightness profile.

In the $M_{\rm BH}$--$\rho_{\rm soi, sph}$ diagram, seven galaxies (NGC~404, IC~2560, NGC~3079, NGC~4388, NGC~4826, NGC~5055, NGC~6323) are considerably offset (from the main population) towards low $M_{\rm BH}$ and $\rho_{\rm soi, sph}$.  Another (S\'ersic) galaxy,  NGC~0821, appears somewhat offset towards a high $\rho_{\rm soi, sph}$ for its black hole mass.  

NGC~404, which is the only galaxy with an intermediate-mass black hole (IMBH) in our sample, is a genuine outlier, and it is possible that IMBHs may not follow the log-linear $M_{\rm BH}$--(stellar density) scaling relations defined by SMBH hosts. 
It has already been excluded from our correlations (as mentioned in Section \ref{Data}).  
NGC~5055 has an unusually small central stellar velocity dispersion relative to its black hole mass\footnote{It is also possible that the value of $M_{\rm BH}$ in NGC~5055, measured using gas dynamical modeling, may be an overestimate.} \citep[see the $M_{\rm BH}$--$\sigma$ diagram in][their figure 2]{Sahu:2019:b}, resulting in a large $r_{\rm soi}$ and thus a small $\rho_{\rm soi, sph}$.  

Galaxies IC~2560, NGC~3079, NGC~4388, NGC~4826, and NGC~6323 have spheroid S\'ersic indices between 0.58 and 1.15; thus, they have a shallow inner density profile and a small $\rho_{\rm soi,sph}$. 
The S\'ersic galaxy NGC~0821,  on the other hand, has a S\'ersic index of 6.1 and hence, a steep inner density profile and a high $\rho_{\rm soi,sph}$.  It also contains a faint edge-on intermediate-scale disk \citep{Savorgnan:Graham:2016:I}, suggestive of an accretion event. 


Including the above eight galaxies significantly biases the best-fit relation defined by most of the sample; hence, we have excluded these galaxies (plus the Milky way) from our $M_{\rm BH}$--$\rho_{\rm soi, sph}$ relations. 
Here, we do not exclude the stripped galaxies NGC~4342 and NGC~4486B (described in Section \ref{Data}) because we are dealing with the spheroid spatial density at a central radius, which may not be affected by the outer mass stripping of these galaxies.

Initially, we performed a single regression between $M_{\rm BH}$ and $\rho_{\rm soi, sph}$  using our  (S\'ersic  + core-S\'ersic) sample\footnote{\label{single_Mbh_Rho_soi} The single regression provides the relation $\log \left( M_{\rm BH}/{\rm M_{\sun}} \right )=  (-1.27 \pm 0.07) \, \log \left( \rm \rho_{\rm soi, sph}/10^{2.5} \, M_{\sun} pc^{-3} \right)+(8.55 \pm 0.07)$, with $\Delta_{\rm rms|BH}=0.77$ dex.}, as shown in the top panel of Figure \ref{Mbh-Rho_soi}. 
We noticed that the distribution of core-S\'ersic galaxies traces a substructure systematically offset from the best-fit line for the ensemble of galaxies, suggesting a different trend for this sub-sample. 
Therefore, we further performed different regressions for the core-S\'ersic and S\'ersic galaxies, presented in the middle panel of Figure \ref{Mbh-Rho_soi}.
We observed a  tight,  shallower relation for the core-S\'ersic galaxies, given by
\begin{IEEEeqnarray}{rCl}
\label{Mbh_rho_soi_CS}
\rm \log \left( \frac{M_{\rm BH}}{\rm M_\odot } \right) &=& \rm (-0.68 \pm 0.06) \log \left(\frac{ \rho_{\rm soi, sph}}{\rm 10^{2.5} \, M_\odot pc^{-3}} \right)  \nonumber \\
&& +\> (9.06 \pm 0.05), 
\end{IEEEeqnarray}
with $\Delta_{\rm rms|BH}=0.21$ dex. Curiously, this relation has the lowest total rms scatter of all the 
black hole scaling relations\footnote{It is noted that $r_{\rm soi}$ is derived from $M_{\rm BH}$ (Equation \ref{r_soi}). For a roughly similar $\sigma$ among core-S\'ersic galaxies, those with bigger $M_{\rm BH}$ have a larger $r_{\rm soi}$ (see the left-hand panel of Figure \ref{Rho_profiles}). The slope in Equation \ref{Mbh_rho_soi_CS} tracks the average slope across 20-1000 pc of the high-$n$ (red) profiles in Figure \ref{Rho_profiles}.}. 
For S\'ersic galaxies (with $\rm n \gtrsim 1$), we found a relatively steeper relation, 
\begin{IEEEeqnarray}{rCl}
\label{Mbh_rho_soi_S}
\rm \log \left( \frac{M_{\rm BH}}{\rm M_\odot } \right) &=& \rm (-1.18 \pm 0.10) \log\left(\frac{ \rho_{\rm soi, sph}}{\rm 10^{2.5} \, M_\odot pc^{-3}} \right) \nonumber \\
&& +\> (8.39 \pm 0.10), 
\end{IEEEeqnarray}
with $\Delta_{\rm rms|BH}=0.77$ dex. The correlation coefficients for the above two relations are presented in Table \ref{fit parameters2}. 
Here, we needed to know $M_{\rm BH}$ in advance to measure $r_{\rm soi}$ and thus  $\rho_{\rm soi, sph}$, voiding Equations \ref{Mbh_rho_soi_CS} and \ref{Mbh_rho_soi_S} as black hole mass predictor tools but leaving them as constraints for simulations and to predict $\rho_{\rm soi, sph}$ for a given $M_{\rm BH}$ when the host spheroid surface brightness parameters are not known \citep[as done in][using other black hole scaling relations]{Biava2019}.

The scatter in the above relations is smaller than that about  the  $M_{\rm BH}$--$\mu_{\rm 0, sph}$ (and $M_{\rm BH}$--$\Sigma_{\rm 0, sph}$) relations,  indicating that  $M_{\rm BH}$ has a better relation with $\rho_{\rm soi, sph}$,  supporting the prediction in \citet{Graham:Driver:2007}. 
On their own, the core-S\'ersic galaxies appear to have no correlation in the $M_{\rm BH}$--$\mu_{\rm 0, sph}$ diagram (Figure \ref{Mbh_I0_Ie}). However, the overlapping nature of  (the S\'ersic component of)  their density profiles in the left-hand panel of Figure \ref{Rho_profiles} \citep[also see][their figure 18]{Dullo:Graham:2012}, coupled with Equation \ref{r_soi}, supports the tight trend for cored galaxies seen in Figure \ref{Mbh-Rho_soi}. 
The smaller scatter observed for the core-S\'ersic relation can be understood from the tight distribution of black points marking  $r_{\rm soi}$ and $\rho_{\rm soi, sph}$ on the density profiles of the core-S\'ersic spheroids in the left-hand panel of Figure \ref{Rho_profiles}. 
The green points, marking  $r_{\rm soi}$ and $\rho_{\rm soi, sph}$ on the density profiles of the S\'ersic spheroids are more scattered, explaining the higher rms scatter about the $M_{\rm BH}$--$\rho_{\rm soi, sph}$ relation for the S\'ersic galaxies.

Figure \ref{Rho_profiles} (left-hand panel) also explains why there will be a correlation between black hole mass and the isophotal or isodensity radius measured at faint/low densities. 
It is easy to see that the use of ever-lower densities will result in an ever greater separation of the curves. 
A result due to the different S\'ersic indices ($n$) and the trend between  $M_{\rm BH}$ and $n$  \citep[e.g.,][]{Graham:Driver:2007, Sahu:2020}.

As with the $M_{\rm BH}$--$\rho_{\rm soi, sph}$, a  similar apparent separation between core-S\'ersic and S\'ersic galaxies is recovered in the $M_{\rm BH}$--$\langle \rho \rangle_{\rm soi, sph}$ diagram involving the average spatial density within $r_{\rm soi}$, as shown in the bottom panel of Figure \ref{Mbh-Rho_soi} (see Table \ref{fit parameters2} for fit parameters). However, the scatter is a bit higher than about the $M_{\rm BH}$--$\rho_{\rm soi, sph}$ relations. Again, it should be noted that the values of $\langle \rho \rangle_{\rm soi, sph}$  (and $\rho_{\rm soi, sph}$) for the core-S\'ersic spheroids are higher than the actual values because these are based on the de-projection of the (inwardly extrapolated) S\'ersic portion of their surface brightness profiles, which intentionally do not account for the deficit of light (see footnote \ref{def_core}) in the core, $r \lesssim R_{\rm b}$.  

For the core-S\'ersic galaxies, the $M_{\rm BH} \propto$ (stellar  mass deficit: $M_{*,def}^{0.27}$) relation \citep[][their equation 18]{Dullo:2014} suggests that galaxies with high $M_{\rm BH}$ have a higher mass deficit. 
Upon accounting for the  mass deficit to obtain the actual $\rho_{\rm soi, sph, core}$ (and $\langle \rho \rangle_{\rm soi, sph, core}$),  all the core-S\'ersic galaxies will move towards a lower $\rho_{\rm soi, sph}$ (and $\langle \rho \rangle_{\rm soi, sph}$), i.e., towards right-side in Figure \ref{Mbh-Rho_soi} (where the horizontal axes are inverted). 
However, galaxies with higher $M_{\rm BH}$ shall shift more than the galaxies with lower $M_{\rm BH}$, generating a slightly shallower (negative/declining) slope than the slope of the relation presented here (Equation \ref{Mbh_rho_soi_CS}), but still preserving the apparent core-S\'ersic versus S\'ersic substructuring.

The negative correlations between $M_{\rm BH}$ and $\rho_{\rm soi, sph}$ (and $\langle \rho \rangle_{\rm soi, sph}$) can be visualized from the vertical ordering of blue-to-red shades (i.e., low-to-high $M_{\rm BH}$) of the spheroid density profiles, shown in the middle panel of Figure \ref{Rho_profiles}, with the radial-axis normalized at $r_{\rm soi}$. Broadly speaking, at the influence radius (and any fixed multiple of this radius substantially beyond $r/r_{\rm soi} = \rm 1$),  the stellar density increases while going from the high-$M_{\rm BH}$ (reddish profiles) to  low-$M_{\rm BH}$ (bluer profiles). 
The general (negative) $M_{\rm BH}$--$\rho_{\rm soi, sph}$ trend for our sample arises from massive black holes having larger spheres-of-influence, relative to low-mass black holes, combined with the spheroid's radially declining density profiles.
However, the resultant relations for the core-S\'ersic and S\'ersic galaxies are dependent on the sample selection and, thus, the range of S\'ersic profiles included in each subsample, as discussed in the following subsection.

\subsubsection{Investigating the core-S\'ersic versus S\'ersic substructure} 
One may wonder if the substructures in the  $M_{\rm BH}$--$\rho_{\rm soi, sph}$ (and $M_{\rm BH}$--$\langle \rho \rangle_{\rm soi, sph}$) diagrams seen between core-S\'ersic and S\'ersic galaxies may be related to a similar division observed in the $\rm L$--$\sigma$ and $M_{\rm BH}$--$\sigma$  diagrams \citep[see][]{Davies:1983, Held:Mould:1994, Matkovic:Guzman:2005, Bogdan:2018, Sahu:2019:b}. 
This may be because some of the division seen in the $M_{\rm BH}$--$\rho_{\rm soi, sph}$ and $M_{\rm BH}$--$\langle \rho \rangle_{\rm soi, sph}$ diagram may be influenced by the use of the central stellar velocity dispersion while calculating $r_{\rm soi}$. 
Or conversely, the substructures observed in the $M_{\rm BH}$--$\sigma$ diagram may partly be a reflection of the  $M_{\rm BH}$--$\rho_{\rm soi, sph}$ (or $\langle \rho \rangle_{\rm soi, sph}$) relations, if $\rho_{\rm soi, sph}$ influences $\sigma$.

To test this connection, we tried an alternative estimation of the black hole's influence radius denoted by $r_{\rm soi, 2BH}$.  
The radius $r_{\rm soi, 2BH}$ marks the sphere within which the stellar mass is equivalent to twice the central black hole's mass \citep{Merritt2004}.  
Upon using the  internal density ($\rho_{\rm soi, 2BH, sph}$) calculated at $r_{\rm soi, 2BH}$, we recover the substructure between  core-S\'ersic  and S\'ersic  galaxies in the $M_{\rm BH}$--$\rho_{\rm soi, 2BH, sph}$ diagram\footnote{The core-S\'ersic galaxies follow the relation $\log \left( M_{\rm BH}/{\rm M_{\sun}} \right )=(-0.65 \pm 0.07) \, \log \left( \rm \rho_{\rm soi, 2BH, sph}/10^{2.5} \, M_{\sun} pc^{-3} \right)+(8.73 \pm 0.09)$, and S\'ersic galaxies follow $\log \left( M_{\rm BH}/{\rm M_{\sun}} \right )=(-0.98 \pm 0.07) \, \log \left( \rm \rho_{\rm soi, 2BH, sph}/10^{2.5} \, M_{\sun} pc^{-3} \right)+(7.76 \pm 0.10)$, with $\Delta_{\rm rms|BH}=$ 0.29 dex and 0.87 dex, respectively.} (not shown), albeit with an increased scatter. 
This test demonstrated that the substructuring seen in Figure \ref{Mbh-Rho_soi} is not due to the propagation of $\sigma$ via Equation \ref{r_soi}.

To investigate another scenario underlying the apparent substructures in the $M_{\rm BH}$--$\rho_{\rm soi, sph}$ (and $M_{\rm BH}$--$\langle \rho \rangle_{\rm soi, sph}$) diagrams, we color-coded the data points in the top panel of Figure \ref{Mbh-Rho_soi}  according to their S\'ersic indices. 
This S\'ersic index color map  divides the data in the $M_{\rm BH}$--$\rho_{\rm soi, sph}$ diagram in different diagonal zones,  in a sequential order of $n_{\rm sph}$,  such that one can obtain a set of $M_{\rm BH}$--$\rho_{\rm soi, sph}$ relations applicable for different ranges of $n_{\rm sph}$. 
For example, roughly, we can point out three zones in the top panel of Figure \ref{Mbh-Rho_soi}: the excluded data points near the bottom right of the plot with the smallest S\'ersic indices ($n_{\rm sph} \lesssim 1.5$);  the blue-purple-magenta points with $1.5\lesssim  n_{\rm sph} \lesssim 5$ in the middle, and the red-orange-yellow points with $n_{\rm sph} \gtrsim 5$ in the  upper-left part of the diagram. 
Most of our core-S\'ersic galaxies fall in the third zone, which is why we observe them defining a different $M_{\rm BH}$--$\rho_{\rm soi, sph}$ relation than the majority of the S\'ersic galaxies which fall in the second zone. 
 
The distribution of data-points  in the top panel of Figure \ref{Mbh-Rho_soi}  can be better represented on an $M_{\rm BH}$--$\rho_{\rm soi, sph}$--$n_{\rm sph}$ plane. 
This plane will be investigated in our future exploration of a black hole fundamental plane. 
We note that our calculation of $\rho_{\rm soi, sph}$ depends on $n_{\rm sph}$ and $M_{\rm BH}$, and thus these terms are not independently measured quantities. 
As noted, a high $M_{\rm BH}$, associated with a large  $n_{\rm sph}$ \citep[see][ for the $M_{\rm BH}$--$n_{\rm sph}$ relation]{Sahu:2020}, will generate a large $r_{\rm soi}$ and thus lower $\rho_{\rm soi, sph}$.

\subsection{Spatial Mass Density within 1 kpc: The Spheroid Spatial Compactness $\langle \rho \rangle_{\rm 1 kpc, sph}$}
\label{4.2}

The internal mass density is a better measure of the inner density than the projected column density. Hence, we introduce $\langle \rho \rangle_{\rm 1 kpc, sph}$, the spatial version of the projected spheroid compactness $\langle \Sigma \rangle_{\rm 1 kpc,sph}$ (Section \ref{3.2}), defined as the mean internal stellar mass density within the inner 1 kpc of the spheroids.

\begin{figure*}
\begin{center}
\includegraphics[clip=true,trim= 08mm 04mm 15mm 13mm,width=   0.99\textwidth]{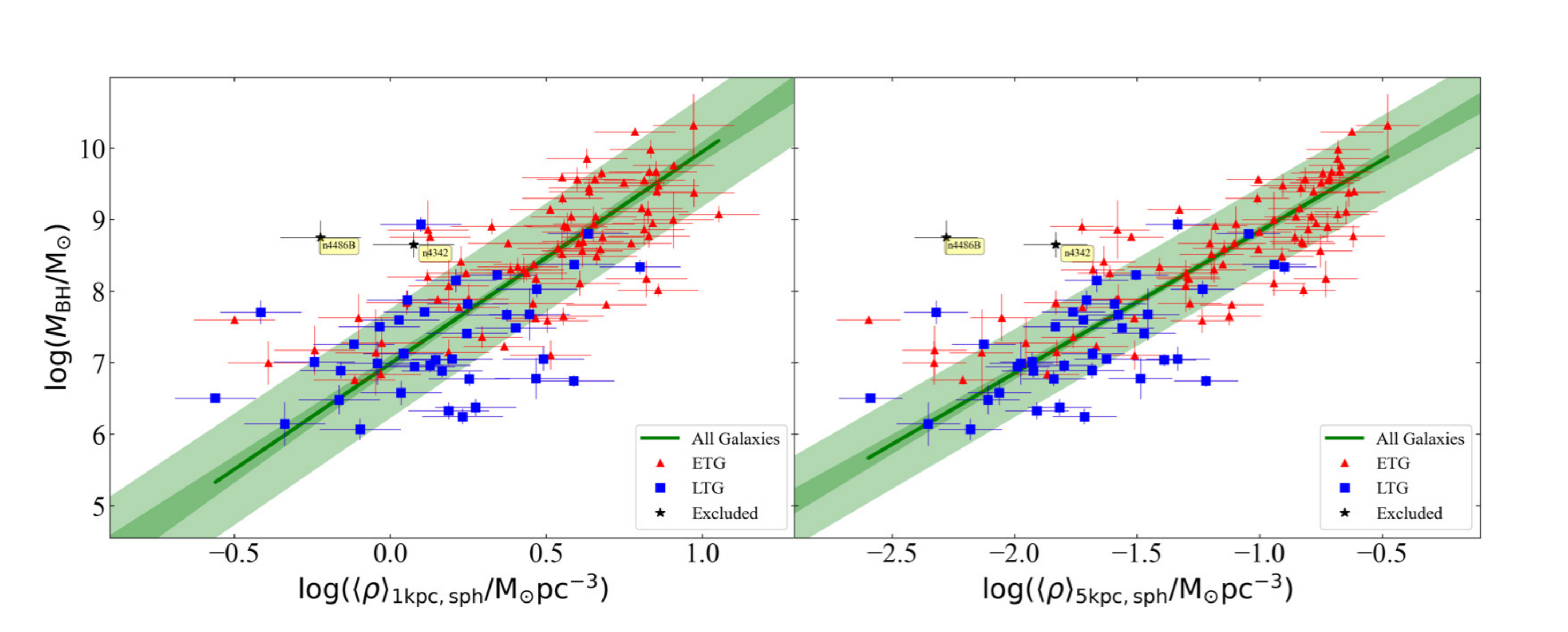}  
\caption{Black hole mass plotted against the internal stellar mass density within the internal spheroid radius of 1 kpc (left-hand panel, Equation \ref{Mbh_rho_1kpc}) and the internal density within the internal spheroid radius of 5 kpc (right-hand panel, Equation \ref{Mbh_rho_5kpc}). Similar to Figure \ref{Compactness}, all galaxy types follow a single relation in these diagrams.}
\label{M-D1}
\end{center}
\end{figure*}

\begin{figure}
\begin{center}
\includegraphics[clip=true,trim= 11mm 05mm 15mm 15mm,width=   0.5\textwidth]{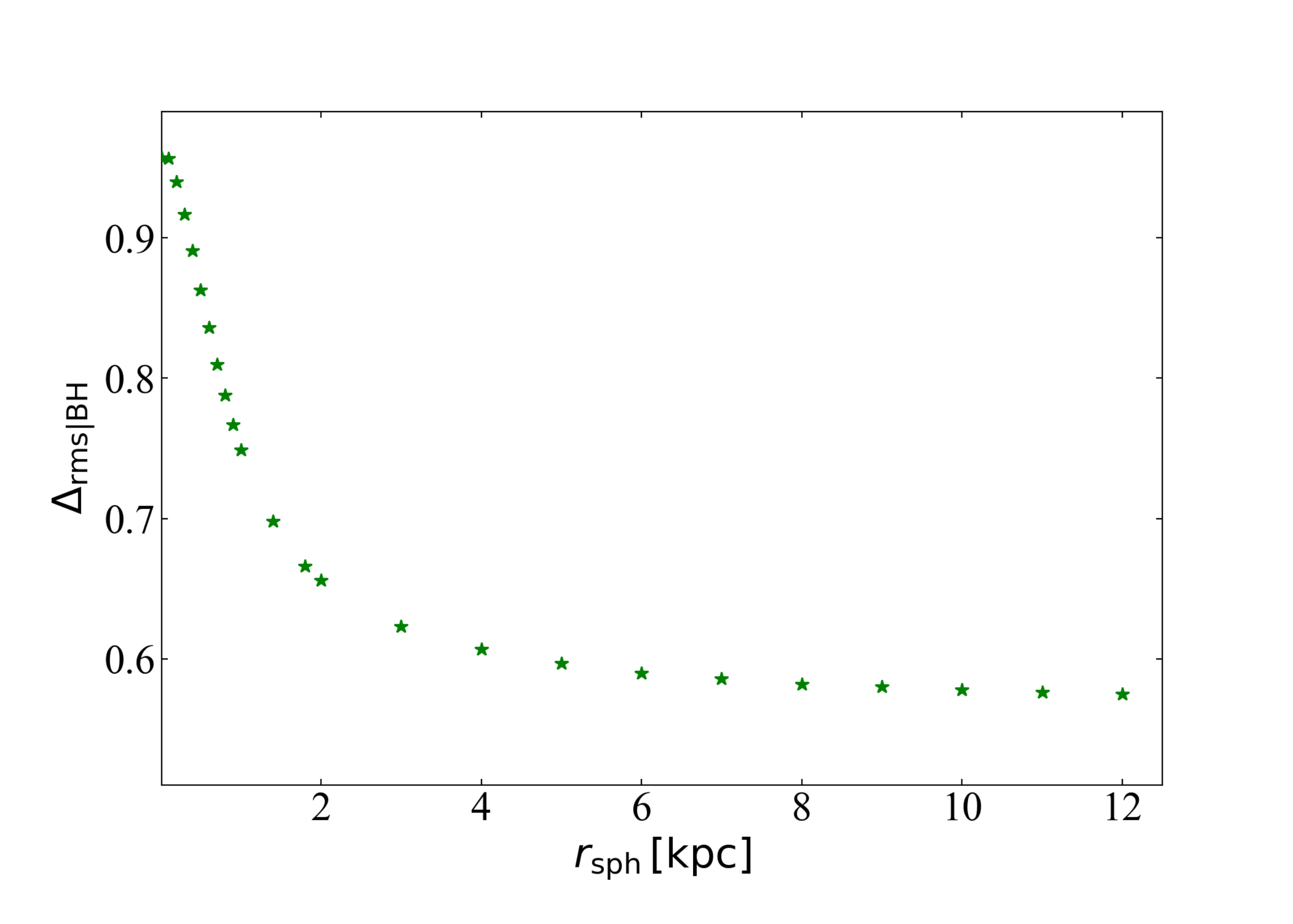}  
\caption{$\Delta_{\rm rms|BH}$ versus $r_{\rm sph}$ for the $M_{\rm BH}$--$\langle \rho \rangle_{\rm r, sph}$ relations.}
\label{Scatter_rho_r}
\end{center}
\end{figure}

We find a positive correlation between the black hole mass and the spheroid spatial compactness without any  detectable substructuring due to the morphological classes of galaxies. 
The single-regression $M_{\rm BH}$--$\langle \rho \rangle_{\rm 1 kpc, sph}$ relation\footnote{Similar to the $M_{\rm BH}$--$\langle \Sigma \rangle_{\rm 1 kpc, sph}$ diagram (Section \ref{3.2}),  we use a $\pm 30\%$ (0.13 dex) uncertainty on  $\langle \rho \rangle_{\rm 1 kpc,sph}$  (and $\langle \rho \rangle_{\rm 5 kpc,sph}$). 
We obtain consistent relations upon using up to $\pm 40\%$ (0.17 dex) uncertainty.}, shown in the left-hand panel of Figure \ref{M-D1},  can be expressed as 
\begin{IEEEeqnarray}{rCl}
\label{Mbh_rho_1kpc}
\rm  \log \left( \frac{M_{\rm BH}}{\rm M_\odot } \right) &=& \rm (2.96 \pm 0.21) \log \left(\frac{\langle \rho \rangle_{\rm 1 kpc, sph}}{\rm 10^{0.5}\,M_\odot \, pc^{-3}}  \right )  \nonumber \\
&& +\> (8.47 \pm 0.07), 
\end{IEEEeqnarray}
and has $\Delta_{\rm rms|BH}=0.75$ dex. 
The $M_{\rm BH}$--$\langle \rho \rangle_{\rm 1 kpc, sph}$  relation is marginally steeper than the $M_{\rm BH}$--$\langle \Sigma \rangle_{\rm 1 kpc,sph}$ relation (Equation \ref{Mbh_Sigma_1kpc}), and has  a slightly higher vertical scatter. 
However, the orthogonal (perpendicular to the best-fit line) scatter in both the diagrams is comparable ($\sim$0.24 dex). 

We find positive trends between $M_{\rm BH}$ and the internal spheroid density within other constant radii (e.g., 0.1 kpc, 5 kpc, 10 kpc) as well. The left-hand panel in Figure \ref{Rho_profiles} shows that, in general,  the high-$M_{\rm BH}$ profiles reside above the low-$M_{\rm BH}$ profiles at all radii; thus, the galactic spheroids with higher $M_{\rm BH}$ are relatively denser than the spheroids with lower $M_{\rm BH}$, when compared at a fixed physical radius. 
This partly explains the positive trends obtained for the correlations of black hole mass with the spatial compactness, $\langle \rho \rangle_{\rm 1 kpc, sph}$, and the internal density at/within any fixed spatial radii. 
However, there is a varying scatter in the relations that decreases with larger radii. 

A plot of the vertical $\Delta_{\rm rms|BH}$ versus $r_{\rm sph}$ for the $M_{\rm BH}$--$\langle \rho \rangle_{\rm r, sph}$ relations is shown in Figure \ref{Scatter_rho_r}. 
For $\rm r_{\rm sph} < 1\, kpc$,  the $M_{\rm BH}$--$\langle \rho \rangle_{\rm r, sph}$ relations have a higher scatter than Equation \ref{Mbh_rho_1kpc}, whereas, for  $\rm r_{\rm sph} > 1\, kpc$ the $M_{\rm BH}$--$\langle \rho \rangle_{\rm r, sph}$ relations are relatively stronger and have a gradually decreasing  scatter with increasing $r_{\rm sph}$, analogous to the $M_{\rm BH}$--$\Sigma_{\rm R,sph}$ relations (Section \ref{3.2}). 
This can be readily understood by again looking at the left-hand panel of Figure \ref{Rho_profiles}, even though it shows the density profiles, $\rho$, rather than the somewhat similar mean density profiles, $\langle \rho \rangle$. 
There, one can see a cleaner separation of profiles of different $M_{\rm BH}$ (and S\'ersic index, $n$) when moving to larger radii, which  is due to the increasingly longer tails of the high-$n$ light profiles.

For a comparison, the $M_{\rm BH}$--$\langle \rho \rangle_{\rm 5 kpc, sph}$ relation (see the right-hand panel of Figure \ref{M-D1}), which has $\Delta_{\rm rms|BH}=0.61$ dex, can be expressed as,
\begin{IEEEeqnarray}{rCl}
\label{Mbh_rho_5kpc}
\rm  \log \left( \frac{M_{\rm BH}}{\rm M_\odot } \right) &=& \rm (1.99 \pm 0.11) \log \left(\frac{\langle \rho \rangle_{\rm 5 kpc, sph}}{\rm 10^{-1.5}\,M_\odot \, pc^{-3}}  \right )  \nonumber \\
&& +\> (7.85 \pm 0.06).
\end{IEEEeqnarray}
The smaller scatter in the above relation when compared to the $M_{\rm BH}$--$\langle \rho \rangle_{\rm 1 kpc, sph}$ relation, and the quasi-saturation of $\Delta_{\rm rms|BH}$ for $\rm r_{\rm sph} \gtrsim 5\, kpc$ (Figure \ref{Scatter_rho_r}), suggests that $\langle \rho \rangle_{\rm 5 kpc, sph}$ can be preferred over $\langle \rho \rangle_{\rm 1 kpc, sph}$ to predict $M_{\rm BH}$. 

Overall, the  $M_{\rm BH}$--$\langle \rho \rangle_{\rm r, sph}$ relations are steeper than the $M_{\rm BH}$--$\langle \Sigma \rangle_{\rm R,sph}$ relations for any fixed spheroid radius ($r=R$), with a marginally higher vertical scatter and similar orthogonal scatter. 
Hence, potentially both properties ($\langle \rho \rangle_{\rm r, sph}$ and  $\langle \Sigma \rangle_{\rm R, sph}$) of a spheroid are  equally good predictors of the central black hole's mass. 

\subsection{Internal Density at and within the Spatial Half-Light Radius: $\rho_{\rm e, int, sph} \, \& \, \langle \rho \rangle_{\rm e,int, sph}$}
\label{4.3}

Using the spheroid internal density profiles,  we calculated the spheroid spatial half-mass radius, $r_{\rm e, sph}$, which represents a sphere enclosing $50\%$ of the total spheroid mass (or luminosity, for a constant mass-to-light ratio).  The ratio  $r_{\rm e, sph}/R_{\rm e, sph}$ is approximately 1.33 \citep{Ciotti:1991}.

\begin{figure*}
\begin{center}
\includegraphics[clip=true,trim= 05mm 00mm 10mm 11mm,width=   1.0\textwidth]{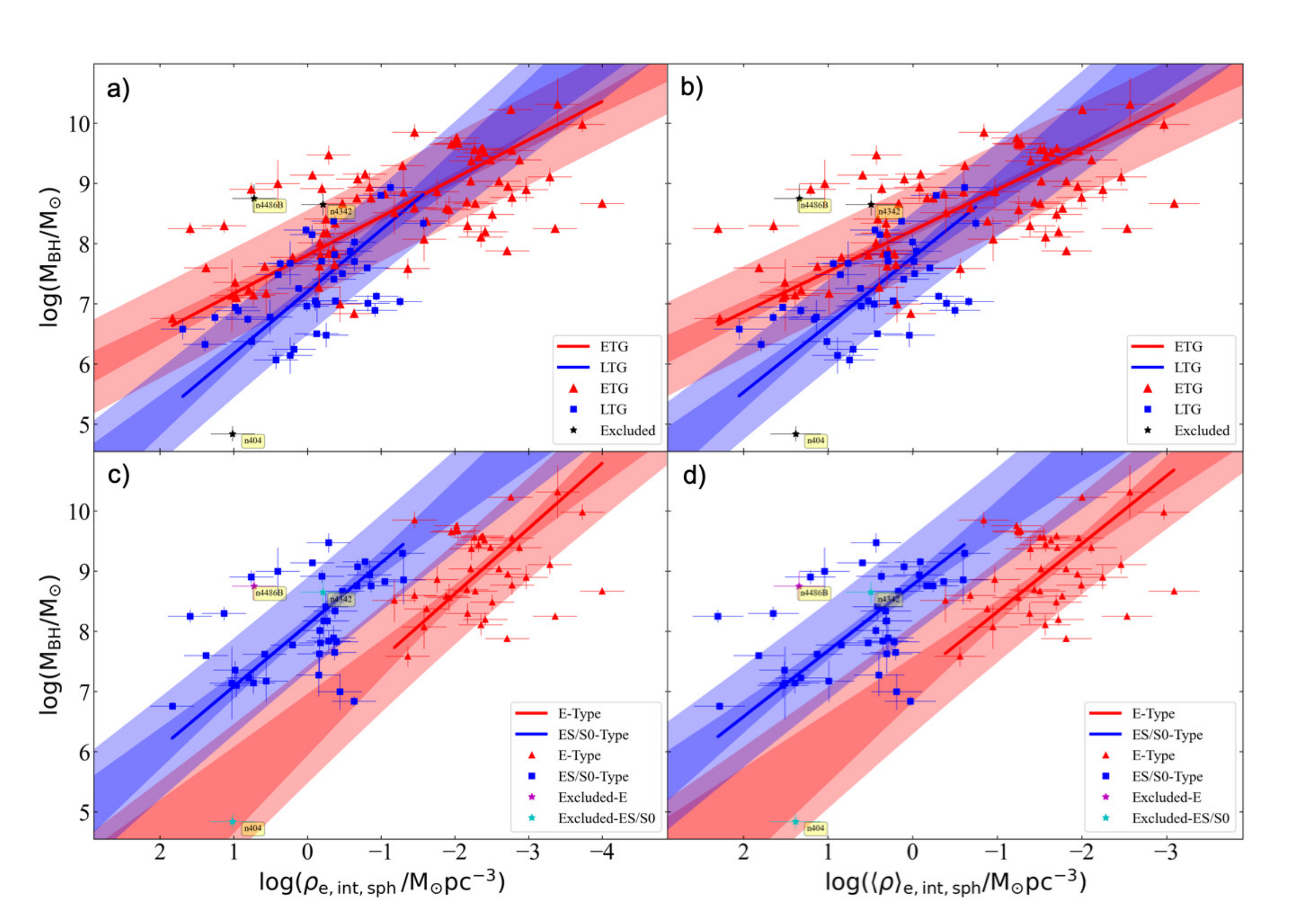}
\caption{Black hole mass versus internal stellar mass density at $r=r_{\rm e,sph}$ (left-hand panels) and within $r_{\rm e,sph}$ (right-hand panels). 
Top panels show that  ETGs and LTGs follow two different $M_{\rm BH}$--$\rho_{\rm e, int, sph}$ relations (panel a, Equations \ref{Mbh_rhoe_ETGs} and \ref{Mbh_rhoe_LTGs}) and $M_{\rm BH}$--$\langle \rho \rangle_{\rm e, sph}$ relations (panel b, Table \ref{fit parameters2}). 
The bottom panels present only ETGs, where ETGs with a  disk (ES- and S0-types) and ETGs without a disk (E-type) are found to follow almost parallel $M_{\rm BH}$--$\rho_{\rm e, int, sph}$ (panel c) and $M_{\rm BH}$--$\langle \rho \rangle_{\rm e, sph}$ (panel d) relations, offset in the vertical direction by more than an order of magnitude (see Table \ref{fit parameters2} for best-fit parameters). 
Note that the horizontal axes of all the panels are inverted, such that the internal density decreases when going from left to right.
}
\label{Mbh-Rho_re}
\end{center}
\end{figure*}

We find that ETGs and LTGs define different (negative) trends between $M_{\rm BH}$ and the internal stellar mass density ($\rho_{\rm e, int, sph}$)  at $r=r_{\rm e, sph}$, as shown in panel-a of Figure \ref{Mbh-Rho_re}. The $M_{\rm BH}$--$\rho_{\rm e, int, sph}$ relation followed by ETGs can be expressed as
\begin{IEEEeqnarray}{rCl}
\label{Mbh_rhoe_ETGs}
\rm  \log \left( \frac{M_{\rm BH}}{\rm M_\odot } \right) &=& \rm (-0.64 \pm 0.04) \log \left( \frac{\rho_{\rm e, int, sph}}{\rm M_\odot pc^{-3}} \right)  \nonumber \\
&& +\> (7.81 \pm 0.10), 
\end{IEEEeqnarray}
with $\Delta_{\rm rms|BH}=0.73$ dex. The steeper relation followed by LTGs, with $\Delta_{\rm rms|BH}=0.69$ dex, is given by 
\begin{IEEEeqnarray}{rCl}
\label{Mbh_rhoe_LTGs}
\rm  \log \left( \frac{M_{\rm BH}}{\rm M_\odot } \right) &=& \rm (-1.02 \pm 0.13) \log \left( \frac{\rho_{\rm e, int, sph}}{\rm M_\odot pc^{-3}}  \right)  \nonumber \\
&& +\> (7.20 \pm 0.11 ).
\end{IEEEeqnarray}
These two relations have a smaller scatter than the $M_{\rm BH}$--$\Sigma_{\rm e, sph}$ relations for ETGs and LTGs (Table \ref{fit parameters1}). The relatively smaller scatter and smaller uncertainties on the fit parameters suggests that $\rho_{\rm e, int, sph}$  can be a better predictor of $M_{\rm BH}$ than $\Sigma_{\rm e, sph}$ (see Table \ref{fit parameters1}).

As we have repeatedly found, the shallower slope for the ETGs is physically meaningless. Its value reflects the sample selection and thus the relative number of ETGs with and without a disk.
Further analysis of the $M_{\rm BH}$--$\rho_{\rm e, int, sph}$ diagram reveals an offset between the ETGs with a rotating stellar disk (ES, S0) and  ETGs without a rotating stellar disk (E), as shown in panel-c of Figure \ref{Mbh-Rho_re}. 
The parameters for the $M_{\rm BH}$--$\rho_{\rm e, int, sph}$ relations obtained for the two ETGs sub-populations are presented in Table \ref{fit parameters2}. 
Notably, these two sub-categories of ETGs follow steeper $M_{\rm BH}$--$\rho_{\rm e, int, sph}$ relations than Equation \ref{Mbh_rhoe_ETGs}, almost parallel to each other but offset from each other by more than an order of magnitude in the $M_{\rm BH}$-direction. 
This offset is analogous to the offset found in the $M_{\rm BH}$--$M_{\rm *, sph}$ \citep{Sahu:2019:a}, $M_{\rm BH}$--$R_{\rm e,sph}$ \citep{Sahu:2020}, and $M_{\rm BH}$--$\langle \Sigma \rangle_{\rm e, sph}$ diagrams (Section \ref{3.3}). 
This offset originates from the smaller effective sizes ($R_{\rm e,sph}$) and higher $\langle \Sigma \rangle_{\rm e, sph}$ of the  ES/S0-type galaxies relative to that of E-type galaxies possibly built from major mergers.

Similar trends and morphological substructures are found between $M_{\rm BH}$ and the average internal density, $\langle \rho \rangle_{\rm e, int, sph}$, within $r_{\rm e,sph}$ (see panels b and d of Figure \ref{Mbh-Rho_re}). 
The parameters for the $M_{\rm BH}$--$\langle \rho \rangle_{\rm e, int, sph}$ relations are  provided in Table \ref{fit parameters2}.
The right-hand panel in Figure \ref{Rho_profiles} presents the spheroid spatial density profiles for our sample with the radial-axis normalized at $r_{\rm e,sph}$. 
At the spatial half-light radius, where $\log (r/r_{\rm e,sph}) = 0$, the increasing spatial density when going from high-$M_{\rm BH}$ to low-$M_{\rm BH}$ profiles is quite clear. 
This explains the negative $M_{\rm BH}$--$\rho_{\rm e, int, sph}$ (and $\langle \rho \rangle_{\rm e, int, sph}$) correlations.  
 
Table \ref{fit parameters2} also provides the morphology-dependent relations obtained between $M_{\rm BH}$ and the spatial density $\langle \rho \rangle_{\rm e, sph}$ within the projected half-light radius ($r=R_{\rm e,sph}$), which are analogous to the substructures in the  $M_{\rm BH}$--$\rho_{\rm e, int, sph}$ diagram. 
The $M_{\rm BH}$--$\langle \rho \rangle_{\rm e, sph}$ relation defined by our ETGs is consistent with that of \citet{Saglia:2016}. 
However, they do not report any of the vital substructures in this diagram due to ETGs (E, ES/S0) and LTGs. 
Without this awareness of the host galaxy morphology, the slope and intercept of the $M_{\rm BH}$--$\langle \rho \rangle_{\rm e, sph}$ relation is meaningless because it is biased by the randomness of one's sample selection. 
Indeed, this is why our ETGs relation has a slope of -0.64  rather than roughly $\rm \sim -1.1$, as followed by the E-type galaxies, the ES/S0-type galaxies, and the spiral galaxies (see Table \ref{fit parameters2}).

Finally, we again note here that similar to the $M_{\rm BH}$--$\Sigma_{\rm e, sph}$ (and $\langle \Sigma \rangle_{\rm e, sph}$) relations (see Figures \ref{Mbh_Ie} and \ref{Mbh_Ie_E_ESS0} in Section \ref{3.3}), the complete  picture of the $M_{\rm BH}$--$\rho_{\rm e, int, sph}$ (and $\langle \rho \rangle_{\rm e,int, sph}$) distributions are curved, which may be revealed in future using a larger sample. The slopes of the linear relations presented here are dependent on the mass range of our sample.

\begin{deluxetable*}{lcllccc}
\tabletypesize{\footnotesize}
\tablecolumns{7}
\tablecaption{Correlations between the Black Hole Mass and the Spheroid Internal Density \label{fit parameters2}}
\tablehead{
\colhead{ \textbf{Category} } & \colhead{ \textbf{Number} } & \colhead{ \bm{$\log ( M_{\rm BH}/ M_{\sun} )=(Slope)\, X + Intercept$} } & \colhead{ \bm{$\epsilon$} }  & \colhead{ \bm{ $\Delta_{\rm rms | BH}$} }  & \colhead{ \bm{$r_p$ } } & \colhead{ \bm{$r_s$}}  \\
\colhead{} & \colhead{} &  \colhead{\textbf{dex}} & \colhead{\textbf{dex}} & \colhead{\textbf{dex}} & \colhead{}  & \colhead{}  \\
\colhead{ \textbf{(1)}} & \colhead{\textbf{(2)}} & \colhead{\textbf{(3)}} & \colhead{\textbf{(4)}} & \colhead{\textbf{(5)}} & \colhead{\textbf{(6)}} & \colhead{\textbf{(7)}} 
}
\startdata
& & \textbf{Internal Density at $r_{\rm soi}$ (Figure \ref{Mbh-Rho_soi}, middle panel)}  & & & & \\
\hline
\vspace*{1mm}
Core-S\'ersic & 31  & $\log \left( M_{\rm BH}/{\rm M_{\sun}} \right )=  (-0.68 \pm 0.06) \, \log \left( \rm \rho_{\rm soi, sph}/10^{2.5} \, M_{\sun} pc^{-3} \right)+(9.06 \pm 0.05)$ & 0.00  & 0.21 &  -0.92   & -0.93 \\ 
\vspace*{1mm}
S\'ersic & 83\tablenotemark{a}  & $\log \left(M_{\rm BH}/{\rm M_{\sun}} \right )=  (-1.18 \pm 0.10) \, \log \left( \rm \rho_{\rm soi, sph}/10^{2.5} \, M_{\sun} pc^{-3} \right)+(8.39 \pm 0.10)$ & 0.68  & 0.77 &  -0.54  &   -0.50 \\ 
\hline
& & \textbf{Internal Density within $r_{\rm soi}$ (Figure \ref{Mbh-Rho_soi}, bottom panel) }  & & & & \\
\hline
\vspace*{1mm}
Core-S\'ersic & 31  & $\log \left( M_{\rm BH}/{\rm M_{\sun}} \right )=  (-0.69 \pm 0.07) \, \log \left( \rm \langle \rho \rangle_{\rm soi, sph}/10^{2.5} \, M_{\sun} pc^{-3} \right)+(9.29 \pm 0.04)$ & 0.07  & 0.25 &  -0.89   & -0.92 \\ 
\vspace*{1mm}
S\'ersic & 83\tablenotemark{a}  & $\log \left(M_{\rm BH}/{\rm M_{\sun}} \right )=  (-1.14 \pm 0.09) \, \log \left( \rm \langle \rho \rangle_{\rm soi, sph}/10^{2.5} M_{\sun} pc^{-3} \right)+(8.59 \pm 0.12)$ & 0.78  & 0.85 &  -0.43  &   -0.40 \\ 
\hline
& & \textbf{Internal Density within 1 kpc of Spheroid (Figure \ref{M-D1}, left-hand panel) }  & & & & \\
\hline
\vspace*{1mm}
All Galaxies & 119  & $\log \left(M_{\rm BH}/{\rm M_{\sun}}  \right )=  (2.96 \pm 0.21) \, \log \left( \rm \langle \rho \rangle_{1 kpc,sph}/ 10^{0.5}\, M_{\sun} pc^{-3} \right)+(8.47 \pm 0.07)$ & 0.63  & 0.75 &  0.73   & 0.75 \\ 
\hline
& & \textbf{Internal Density within 5 kpc of Spheroid (Figure \ref{M-D1}, right-hand panel)}   & & & & \\
\hline
\vspace*{1mm}
All Galaxies & 119  & $\log \left(M_{\rm BH}/{\rm M_{\sun}}  \right )=  (1.99 \pm 0.11) \, \log \left( \rm \langle \rho \rangle_{5 kpc,sph}/10^{-1.5} \, M_{\sun} pc^{-3} \right)+(7.85 \pm 0.06)$ & 0.53  & 0.61 &  0.82  & 0.84 \\ 
\hline
& & \textbf{Internal Density at $r_{\rm e, sph}$ (Figure \ref{Mbh-Rho_re}, left-hand panels)}  & & & & \\
\hline
\vspace*{1mm}
LTGs & 39  & $\log \left(M_{\rm BH}/{\rm M_{\sun}} \right )=  (-1.02 \pm 0.13) \, \log \left( \rm \rho_{e,int,sph}/M_{\sun} pc^{-3} \right)+(7.20 \pm 0.11)$ & 0.63 & 0.69 & -0.56   & -0.58 \\ 
\vspace*{1mm}
ETGs (E+ES/S0) & 80  & $\log \left(M_{\rm BH}/{\rm M_{\sun}} \right )=  (-0.64 \pm 0.04) \, \log \left( \rm \rho_{e,int,sph}/M_{\sun} pc^{-3} \right)+(7.81 \pm 0.10)$ & 0.70 & 0.73 &  -0.64  & -0.61 \\ 
\vspace*{1mm}
E & 40  & $\log \left(M_{\rm BH}/{\rm M_{\sun}}  \right )=  (-1.09 \pm 0.11) \, \log \left( \rm \rho_{e,int,sph}/M_{\sun} pc^{-3} \right)+(6.45 \pm 0.26)$ & 0.76  & 0.82 &  -0.26    & -0.21 \\ 
\vspace*{1mm}
ES/S0 & 40  & $\log \left(M_{\rm BH}/{\rm M_{\sun}}  \right )=  (-1.03 \pm 0.10) \, \log \left( \rm \rho_{e,int,sph}/M_{\sun} pc^{-3} \right)+(8.11 \pm 0.12)$ & 0.70  & 0.77 &  -0.52   & -0.50 \\ 
\hline
& & \textbf{Internal Density within $r_{\rm e, sph}$ (Figure \ref{Mbh-Rho_re}, right-hand panels)}  & & & & \\
\hline
\vspace*{1mm}
LTGs & 39  & $\log \left(M_{\rm BH}/{\rm M_{\sun}} \right )=  (-1.12 \pm 0.15) \, \log \left(\rm  \langle \rho \rangle_{\rm e,int,sph}/M_{\sun} pc^{-3} \right)+(7.78 \pm 0.13)$ & 0.64 & 0.72 & -0.53   & -0.56 \\ 
\vspace*{1mm}
ETGs (E+ES/S0) & 80  & $\log \left(M_{\rm BH}/{\rm M_{\sun}} \right )=  (-0.68 \pm 0.05) \, \log \left(\rm  \langle \rho \rangle_{\rm e,int,sph}/M_{\sun} pc^{-3} \right)+(8.22 \pm 0.09)$ & 0.70 & 0.74 &  -0.63  & -0.59 \\ 
\vspace*{1mm}
E & 40  & $\log \left(M_{\rm BH}/{\rm M_{\sun}}  \right )=  (-1.13 \pm 0.13) \, \log \left(\rm  \langle \rho \rangle_{\rm e,int,sph}/M_{\sun} pc^{-3}  \right)+(7.20 \pm 0.21)$ & 0.75  & 0.81 &  -0.26    & -0.17 \\ 
\vspace*{1mm}
ES/S0 & 40  & $\log \left(M_{\rm BH}/{\rm M_{\sun}}  \right )=  (-1.09 \pm 0.12) \, \log \left( \rm \langle \rho \rangle_{\rm e,int,sph}/M_{\sun} pc^{-3}  \right)+(8.77 \pm 0.12)$ & 0.71  & 0.79 &  -0.49   & -0.44 \\ 
\hline
& & \textbf{Internal Density within $R_{\rm e, sph}$}  & & & & \\
\hline
\vspace*{1mm}
LTGs & 39  & $\log \left(M_{\rm BH}/{\rm M_{\sun}} \right )=  (-1.16 \pm 0.16) \, \log \left(\rm  \langle \rho \rangle_{\rm e,sph}/M_{\sun} pc^{-3} \right)+(8.04 \pm 0.16)$ & 0.65 & 0.73 & -0.52   & -0.54 \\ 
\vspace*{1mm}
ETGs (E+ES/S0) & 80  & $\log \left(M_{\rm BH}/{\rm M_{\sun}} \right )=  (-0.68 \pm 0.05) \, \log \left(\rm  \langle \rho \rangle_{\rm e,sph}/M_{\sun} pc^{-3} \right)+(8.40 \pm 0.08)$ & 0.70 & 0.74 &  -0.63  & -0.59 \\ 
\vspace*{1mm}
E & 40  & $\log \left(M_{\rm BH}/{\rm M_{\sun}}  \right )=  (-1.13 \pm 0.13) \, \log \left(\rm  \langle \rho \rangle_{\rm e,sph}/M_{\sun} pc^{-3}  \right)+(7.50 \pm 0.19)$ & 0.75  & 0.81 &  -0.26    & -0.17 \\ 
\vspace*{1mm}
ES/S0 & 40  & $\log \left(M_{\rm BH}/{\rm M_{\sun}}  \right )=  (-1.11 \pm 0.12) \, \log \left( \rm \langle \rho \rangle_{\rm e,sph}/M_{\sun} pc^{-3}  \right)+(9.06 \pm 0.13)$ & 0.71  & 0.79 &  -0.49   & -0.44 \\ 
\enddata
\tablecomments{ Column names are same as in Table \ref{fit parameters1}.
}
\tablenotetext{a}{After excluding eight significant outliers, marked in Figure \ref{Mbh-Rho_soi}, which can significantly affect the best-fit line for the ensemble of S\'ersic galaxies.}
\end{deluxetable*}

\section{Implications and Discussion}
\label{Discussion}
 
\subsection{Prediction of $M_{\rm BH}$} 
\label{5.1}
We have shown how and explained why the BH mass correlates with a range of projected and internal stellar densities of the host spheroid. Plotting the density profiles of the (123 $-$ Milky Way=) 122 spheroids together in the same figure reveals that the spheroids with larger BH masses reside in profiles with larger half-light radii, higher S\'ersic indices, and longer tails to the light profile (Figures \ref{Sersic_all} and \ref{Rho_profiles}). 
At larger radii, the separation of spheroids with low-$M_{\rm BH}$ and low-$n$  profiles from those with high-$M_{\rm BH}$ and high-$n$ profiles becomes cleaner. 
Consequently, and counter-intuitively, the use of densities calculated at larger radii yields less scatter in the  $M_{\rm BH}$--density diagram (Section \ref{3.2} and \ref{4.2}).

The $M_{\rm BH}$--$\langle \Sigma\rangle_{\rm 5 kpc,sph}$ relation (Equation \ref{Mbh_Sigma_5kpc}) and  the $M_{\rm BH}$--$\langle \rho \rangle_{\rm 5 kpc,sph}$ relation (Equation \ref{Mbh_rho_5kpc})  have similar rms scatters (0.59 dex and 0.61 dex), and are applicable to all galaxy types. 
The scatter in these diagrams is comparable to the morphology-dependent $M_{\rm BH}$--$M_{\rm *,sph}$ relations (cf.,  0.50 dex, 0.57 dex, and 0.64 dex for E-, ES/S0-, and S-types, respectively) and the $M_{\rm BH}$--$R_{\rm e,sph}$ relations (cf., 0.59 dex, 0.61 dex, and 0.60 dex for E-, ES/S0-, and S-types, respectively), and smaller than the morphology-dependent $M_{\rm BH}$--$n_{\rm sph}$ relation (cf., 0.73 dex and 0.68 dex for ETGs and LTGs, respectively). 
Thus, $\langle \Sigma\rangle_{\rm 5 kpc,sph}$ and $\langle \rho \rangle_{\rm 5 kpc,sph}$ can predict $M_{\rm BH}$ as good as predicted using $M_{\rm *,sph}$ and $R_{\rm e,sph}$, and  better than $M_{\rm BH}$ predicted using $n_{\rm sph}$.
However, the density at 5 kpc may be very low for spheroids with $R_{\rm e}$ less than half a  kpc, and these relations become more of a reflection of the $M_{\rm BH}$--n relations, and to a lesser degree the $M_{\rm BH}$--$R_{\rm e}$ relations \citep{Sahu:2020}.

The $\rm 3.6 \, \mu m$ $M_{\rm BH}$--$\mu_{\rm 0, sph}$ (Equation \ref{Mbh_mu0})  and the $M_{\rm BH}$--$\mu_{\rm e, sph}$ relations  (see Table \ref{fit parameters1}) offer an alternative way to predict $M_{\rm BH}$ using $\mu_{\rm 0, sph}$ or $\mu_{\rm e, sph}$, just from a calibrated ($\rm 3.6 \, \mu m$)  spheroid surface brightness profile, without requiring either galaxy distance (for local galaxies where the cosmological corrections are very small) or a  stellar mass-to-light ratio which can be complicated to choose. 
However, due to a higher scatter about these relations,  the error bars on the predicted  $M_{\rm BH}$ will be higher than obtained using the $M_{\rm BH}$--$n_{\rm sph}$ and $M_{\rm BH}$--$R_{\rm e, sph}$ relations \citep[see][]{Sahu:2020}. The values $n_{\rm sph}$  and $R_{\rm e, sph}$ can also be obtained from an uncalibrated surface brightness profile. 
Plausibly, the high scatter in the $M_{\rm BH}$--$\mu_{\rm 0, sph}$ diagram is due to the use of a \textit{column} density, and the high scatter in the $M_{\rm BH}$--$\mu_{\rm e, sph}$ diagram arises from a curved distribution of points.

For comparison, the $M_{\rm BH}$--$\phi$ relation for spiral galaxies \citep{Seigar:Kennefick:2008, Berrier:Davis:2013} also has a small scatter of 0.43 dex \citep{Davis:Graham:2017}, where $\phi$ is the pitch angle, i.e., the winding angle of the spiral arms. This relation can provide good estimates of $M_{\rm BH}$ for spiral galaxies.
Including all galaxy types, the $M_{\rm BH}$--$\sigma$  relation has a scatter of 0.53 dex; however, the $M_{\rm BH}$--$\sigma$ diagram has different relations for  core-S\'ersic (cf., 0.46 dex)  and S\'ersic (cf., 0.55 dex) galaxies, which can provide a better estimate of $M_{\rm BH}$ than the single relation, if the core-S\'ersic or S\'ersic morphology is known. 
Another, preferred relation to predict $M_{\rm BH}$ may be the morphology-dependent  $M_{\rm BH}$--$M_{\rm *,gal}$ relation \citep[cf., 0.58 dex and 0.79 dex for ETGs and LTGs, respectively][]{Sahu:2019:a}, where, one does not need to go through the multi-component decomposition process to obtain the  galaxy stellar mass, $M_{\rm *,gal}$. 

The tight $M_{\rm BH}$--$\rho_{\rm soi,sph}$ relation for the core-S\'ersic galaxies has the least total scatter (0.21 dex, see Table \ref{fit parameters2}) among all the black hole scaling relations; whereas the  $M_{\rm BH}$--$\rho_{\rm soi,sph}$ relation obtained for the S\'ersic galaxies has a higher scatter (0.77 dex). 
The relation for core-S\'ersic galaxies only captures the upper envelope of high-$n_{\rm sph}$  spheroids in the $M_{\rm BH}$--$\rho_{\rm soi,sph}$  diagram, while the relation for S\'ersic galaxies describes the average relation for spheroids with a medium value of $n$ (between $\sim 1.5 \,  \rm to \, \sim 5$). 
Overall, the $M_{\rm BH}$--$\rho_{\rm soi,sph}$   diagram suggests that the inclusion of $n_{\rm sph}$ as a third parameter will lead to a black hole plane with a considerably reduced scatter. 
However, if it was to turn out that the mass of the black hole is better connected to the stellar density within its sphere of influence and the stellar concentration (quantified by $n$), it is not useful for predicting  $M_{\rm BH}$, because $\rho_{\rm soi,sph}$ requires knowledge of $r_{\rm soi}$ and thus $M_{\rm BH}$.

\subsection{Dependence of the Black Hole Scaling Relations on the Galaxy Morphology} 
\label{5.2}

\citet{Sahu:2020} did not report on the offset between the ETG subpopulations (E vs ES/S0-types) in the $M_{\rm BH}$--$\mu_{\rm e, sph}$ (or $\langle  \mu \rangle_{\rm e, sph}$, or $\Sigma_{\rm e, sph}$, or $\langle  \Sigma \rangle_{\rm e, sph}$) diagrams, that we reinvestigated here. 
Our investigation here has revealed an offset between the  E- and ES/S0-type galaxy samples (Figure \ref{Mbh_Ie_E_ESS0}). However,  the $M_{\rm BH}$--$\mu_{\rm e, sph}$ correlations obtained for the E- and ES/S0-types are weak, and their slopes and the offset are not established. 
This is plausibly because they follow a curved relation with varying slopes, and we have sampled the bend points of the curves (see Figure \ref{Mbh_Ie_E_ESS0}). Consequently, there is not a strong correlation between $M_{\rm BH}$ and the various effective densities for our sample (Section \ref{3.3}).

Morphology-dependent divisions in the $M_{\rm BH}$--$n_{\rm sph}$ (ETG vs LTG),  $M_{\rm BH}$--$R_{\rm e, sph}$  (E vs ES/S0  vs LTG), and, as seen here, the  $M_{\rm BH}$--$\mu_{\rm e, sph}$  (E vs ES/S0  vs LTG) diagrams,  propagate into the $M_{\rm BH}$--$M_{\rm *,sph}$   \citep[E vs ES/S0 vs LTG,][]{Sahu:2019:a} diagrams. 
Similarly, these morphological substructures are also propagated to the $M_{\rm BH}$--$\rho_{\rm e, int, sph}$ (and $\langle\rho\rangle_{\rm e, int, sph}$) diagrams presented here  (Figure \ref{Mbh-Rho_re}).
Although the ETGs and LTGs seem to define distinct tight relations,  there is an order of magnitude offset\footnote{This offset between ETG with and without a disk is minimized in the $M_{\rm BH}$--$M_{\rm *, gal}$ diagram, where  \citet{Sahu:2019:a} revealed only two distinct relations due to (all) ETGs and LTGs.} in the $M_{\rm BH}$-direction between ETGs without a disk (E-type or slow-rotators) and ETGs with a disk (ES/S0-types or fast-rotators).  
The offset between E- and ES/S0-type galaxies is a combined effect of a smaller bulge size ($R_{\rm e, sph}$) and brighter $\mu_{\rm e, sph}$  (higher $\Sigma_{\rm e, sph}$ and $\langle  \Sigma \rangle_{\rm e, sph}$) of the  ES/S0-type galaxies compared to that of E-type galaxies hosting a similar black hole mass (Section \ref{3.3}). 

As discussed in Section \ref{4.1}, the S\'ersic versus core-S\'ersic division in the $M_{\rm BH}$--$\rho_{\rm soi, sph}$ diagram (Figure \ref{Mbh-Rho_soi}) remains independent of whether or not $r_{\rm soi}$ is calculated using the central stellar velocity dispersion. 
Hence, the S\'ersic versus core-S\'ersic substructures observed in the $M_{\rm BH}$--$\sigma$ diagram \citep{Sahu:2019:b} and the $M_{\rm BH}$--$\rho_{\rm soi, sph}$ (or $\langle\rho\rangle_{\rm soi, sph}$) diagrams are not directly related. 
Nonetheless, the $M_{\rm BH}$--$\sigma$ and $M_{\rm BH}$--$\rho_{\rm soi, sph}$ relations are respectively aware of the galaxy morphology, and thus the galaxies' evolutionary tracks  and their central light/mass concentration, i.e., S\'ersic index (see the top panel in Figure \ref{Mbh-Rho_soi} and the description in Section \ref{4.1}).

\subsection{Fundamental Black Hole Scaling Relation}
\label{5.3}

Many studies have suggested that the $M_{\rm BH}$--$\sigma$ relation may be the most fundamental/universal relation \citep[e.g.,][]{Ferrarese:Merritt:2000, Gebhardt:2000, Ferrarese:Ford:2005, Nicola:2019, Marsden:Shankar:2020} between a black hole and the host galaxy due to its obvious link with the galaxy's gravitational potential and the appearance of $M_{\rm BH} \propto \sigma^{4 - 5}$ relations in theories trying to explain black hole feedback \citep{Silk:Rees:1998, Fabian:1999}. 
These claims are based on  past observations \citep{van_den_Bosch:2016, Saglia:2016} which reported a single $M_{\rm BH}$--$\sigma$ relation for all galaxy types (including bulge-less galaxies), and also a smaller scatter seen in the $M_{\rm BH}$--$\sigma$ diagram relative to the $M_{\rm BH}$--$M_{\rm *,sph}$ relation.
However, over the years, increments in the scatter about the $M_{\rm BH}$--$\sigma$ relation to $\sim0.5$ dex  with growing sample size \citep[see the introduction in][]{Sahu:2019:b}, plus the revelation of a  S\'ersic  ($M_{\rm BH} \propto \sigma^{\sim5}$) versus core-S\'ersic ($M_{\rm BH} \propto \sigma^{\sim 8}$) division in the $M_{\rm BH}$--$\sigma$ diagram \citep[e.g.,][]{Bogdan:2018, Sahu:2019:b, Dullo:GildePaz:2020}, undermine the  perceived superiority of $\sigma$. 
 
Importantly, if the relation with the least scatter should be the primary criteria for deciding the fundamental black hole scaling relation, recent studies further confound the situation.
For example: the $M_{\rm BH}$--$\rho_{\rm soi, sph}$ relation (Equation \ref{Mbh_rho_soi_CS}) for core-S\'ersic galaxies has a total rms scatter of 0.21 dex; the  $M_{\rm BH}$--($R_{\rm b}$: break radius) relation for core-S\'ersic galaxies has $\Delta_{\rm rms|BH} =$ 0.29 dex \citep{Dullo:GildePaz:2020}; the  $M_{\rm BH}$--(pitch angle) relation for spiral galaxies has $\Delta_{\rm rms|BH} =$ 0.43 dex \citep{Davis:Graham:2017}; and  the $M_{\rm BH}$--$M_{\rm *,sph}$ relation for ETGs has $\Delta_{\rm rms|BH} =$ 0.52 dex  \citep{Sahu:2019:a}. 
Moreover, the substructure in the $M_{\rm BH}$--$\rho_{\rm soi, sph}$ diagram (Figure \ref{Mbh-Rho_soi}) due to different ranges of $n_{\rm sph}$ values suggest the existence of a possibly stronger $M_{\rm BH}$--$\rho_{\rm soi, sph}$--$n_{\rm sph}$ plane, which shall be investigated in  future work. Of course, $\rho_{\rm soi, sph}$ is calculated using $M_{\rm BH}$, so some care will be required in such an exploration.

\subsection{Super Massive Black Hole Binary Merger Timescale}
\label{5.4}
The stellar density around a super massive black hole binary (SMBHB) plays an essential role in accelerating the merger of the black holes through dynamical friction \citep{Chandrasekhar:1943, Begelman:1980, ArcaSedda:2014}. 
During a galaxy merger,  dynamical friction pushes the black holes towards the core of the galaxy merger remnant, forming a binary at parsec scales. 
The SMBHB goes through a hydro-dynamical interaction with the surrounding stars (and dust/gas), entering a hardening phase, i.e. when the binding energy of the binary exceeds the average kinetic energy of stars around it \citep{Holley-Bockelmann:2016}. 
The binary then transitions from the hardening to the gravitational wave (GW) emission phase, which eventually drives the binary to merge \citep{Celoria:Oliveri:2018}. 
The major part of a binary lifetime is spent in this transition phase/separation \citep{Sesana:Khan:2015}, the orbital frequency at this transition separation is known as the transition frequency. 
This time period ($\approx$ binary lifetime)  can be estimated using the average stellar density ($\langle \rho \rangle_{\rm soi}$), and stellar velocity dispersion ($\sigma_{\rm soi}$) at the sphere-of-influence of the binary and the binary's orbital eccentricity \citep[e.g.,][their equation 7]{Sesana:Khan:2015}.  
The  transition frequency, which is a part of GW strain model (discussed next),  is also estimated using $\langle \rho \rangle_{\rm soi}$, $\sigma_{\rm soi}$, and eccentricity \citep[][their equation 21]{Chen:Sesana:2017}.

Recently, \citet{Biava2019} estimated the SMBHB  lifetime, as discussed above, using S\'ersic parameters of a remnant-bulge hosting a given (binary) black hole mass. 
They used the $M_{\rm BH}$--$M_{\rm *, sph}$ relation \citep{Savorgnan:2016:Slopes}, the $M_{\rm *, sph}$--$\rm R_{e, sph}$ relation \citep{Dabringhausen:2008}, and the $M_{\rm BH}$--$n_{\rm sph}$ relation \citep{Davis:2018:a}  to obtain the S\'ersic parameters of bulges hosting $10^{5}-10^{8} \, M_\sun$ binary black holes, with the assumption that the merger remnants follow these relations. 
Using these bulge parameters, they applied the \citet{Prugniel1997} density model to obtain $\langle \rho \rangle_{\rm soi}$ to estimate the binary lifetime using the model from \citet[][their equation 7]{Sesana:Khan:2015}.

Now, using our $M_{\rm BH}$--$\langle \rho \rangle_{\rm soi}$ relations obtained here and the $M_{\rm BH}$--$\sigma$ relations \citep[e.g.,][]{Sahu:2019:b}, one can directly obtain the $\langle \rho \rangle_{\rm soi}$ values and  the central $\sigma$, respectively, for a given $M_{\rm BH}$, and using the central $\sigma$ as a proxy for  $\sigma_{\rm soi}$, one can  estimate the typical binary lifetime more directly. 
One can also apply the expression of mean aperture correction for stellar velocity dispersion  \citep[from e.g.,][]{Jorgensen:1995, Cappellari:2006} to  drive $\sigma_{\rm soi}$ using the central $\sigma$ (normalized at aperture size of 0.595 kpc) obtained from our $M_{\rm BH}$--$\sigma$ relation and $r_{\rm soi}$.
Similarly, using the $\langle \rho \rangle_{\rm soi}$ and $\sigma$ values for a given $M_{\rm BH}$ (and some binary eccentricity),  the estimation of the transition frequency can be more straightforward \citep[see][their equation 21]{Chen:Sesana:2017}. 
This way, one would not need to go through various black hole scaling relations for the bulge parameters to obtain $\langle \rho \rangle_{\rm soi}$, using an approximation for $\sigma$, and choosing an approximate density model, e.g., as suggested in \citet{Sesana:Khan:2015} and followed in \citet{Biava2019}.

However, one should note that for galaxies with either a nuclear disk or nuclear star cluster, the $\langle \rho \rangle_{\rm soi}$ will be higher than estimated using the $M_{\rm BH}$--$\langle \rho \rangle_{\rm soi}$ relations for just spheroids. 
Whereas, for core-S\'ersic galaxies, the $\langle \rho \rangle_{\rm soi}$ will be lower than estimated using the $M_{\rm BH}$--$\langle \rho \rangle_{\rm soi}$ relation.

\subsection{Predicting the Gravitational Wave Strain}
\label{5.5}
The long-wavelength gravitational waves (GWs: mHz - nHz), emitted during the SMBHB merger, fall in the detection band of pulsar timing arrays (PTAs: $\micro$Hz - nHz), laser interferometer space antenna \citep[LISA: 0.1 Hz to 0.1 mHz,][]{Amaro-Seoane:Audley:2017}, and other planned space interferometers, such as TianQuin \citep{Luo:Chens:2016}. 
These detectors aim to detect the stochastic GW background (GWB) and individual GWs, which are challenging to predict \citep{Sesana:Vecchio:2009, Mingarelli:Lazio:2017}. 
The detectable amplitude (per unit logarithmic frequency) of perturbations due to the GWB is quantified by the characteristic strain ($\rm h_c$), a typical estimate of which is required for different detectors sensitive to different wavelength ranges of GWs \citep[e.g., see the sensitivity curves for various detectors in][]{Moore:Cole:2015}. 

The GWB characteristic strain can be modeled by integrating the SMBHB merger rate across redshift for a range of chirp-mass\footnote{Chirp mass of a binary comprising of objects with masses $M_{\rm 1}$ and  $M_{\rm 2}$ is given by $\mathcal{M}= (M_{\rm 1} M_{\rm 2})^{3/5}/(M_{\rm 1} +M_{\rm 2})^{1/5}$ \citep[e.g., see][]{Cutler:1994}. It influences the orbital evolution of the binary, e.g., the orbital frequency which governs the emitted GW frequency.} \citep[see the model described in][]{Chen2019}. 
The estimation of SMBHB merger rate is dependent on the observed galaxy mass function, galaxy pair fraction, SMBHB merger time scale (galaxy merger time scale $+$ binary lifetime), and the (black hole)--galaxy scaling relations \citep{Sesana:2013}. 
The (black hole)--galaxy scaling relations convert the galaxy mass function and the galaxy pair fraction into the black hole mass function (BHMF) and the black hole pair fraction (BHPF). 

Often, a constant $M_{\rm *, sph}$/$M_{\rm *, gal}$ ratio has been combined with the old linear $M_{\rm BH}$--$M_{\rm *, sph}$ relation  to obtain an $M_{\rm BH}$/$M_{\rm *, gal}$ ratio, which is used to convert  the galaxy mass function  into the BHMF   \citep[e.g.,][]{Shannon:2015, Chen2019}. 
This causes a bias in the estimated GWB characteristic strain   \citep[e.g.,][show that a quadratic $M_{\rm BH}$--$M_{\rm *, sph}$ relation, instead of a linear relation, changes the predicted extreme mass-ratio inspiral event rate by an order of magnitude]{Mapelli:2012:GW:Space:detectors}. 
The use of our  new morphology-dependent $M_{\rm BH}$--$M_{\rm *, gal}$ relations \citep{Sahu:2019:a} will provide a direct way to obtain a better BHMF and BHPF.  
Coupled with these, the better estimates of the binary lifetime (Section \ref{5.4}) will improve the SMBHB merger rate, which will ultimately improve the predictions for the detectable GWB strain for PTAs and GW space missions. 

\subsection{Tidal Disruption Event Rate}
\label{5.6}
The $M_{\rm BH}$--$\langle \rho \rangle_{\rm soi}$ relation obtained here can also help model the rate of tidal disruption events  \citep[TDEs,][]{Hills:1975}. This is important because apart from probing the black hole population and their environments (especially for BHs in inactive galaxies), TDEs are  used to estimate the black hole mass \citep{Mockler:Guillochon:2019, Zhou:Liu:2021}, and electromagnetic counterparts of the extreme mass-ratio inspirals (EMRIs).  

TDEs are expected to occur more frequently in galaxies with an elevated central stellar density or a nuclear star cluster \citep{Frank:Rees:1976}. The TDEs also require $M_{\rm BH} \lesssim 10^{8} \, \rm M_\odot$ because the weaker tidal forces at, and beyond, the Schwarzchild-Droste radii \citep{Schwarzschild:1916, Droste:1917} of more massive black holes are insufficient to tear open stars and produce a TDE \citep{Rees:1988, Komossa:TDE:2015}.
The TDE rate ($\Gamma_{\rm TDE}$) versus $\langle \rho \rangle_{\rm soi}$ relation in \citet[][their equation 8]{Pfister:Volonteri:2020}  provides a lower limit of $\Gamma_{\rm TDE}$ for a given $\langle \rho \rangle_{\rm soi}$. 
Combining their $\Gamma_{\rm TDE}$--$\langle \rho \rangle_{\rm soi}$ relation with our $M_{\rm BH}$--$\langle \rho \rangle_{\rm soi}$ relation for S\'ersic galaxies, we can  obtain a relation between $M_{\rm BH}$ and TDE rate as
\begin{IEEEeqnarray}{rCl}
\label{TDE}
 \Gamma_{\rm TDE}/\rm year^{-1}= 0.16\times (M_{\rm BH}/M_\odot)^{-0.6},
\end{IEEEeqnarray}
which can be used to obtain a typical estimate of the TDE rate for a given $M_{\rm BH}$. 
This can be refined further through the use of a set of $M_{\rm BH}$--$\langle \rho \rangle_{\rm soi}$ relations, applicable for different ranges of S\'ersic index, or the creation of an $M_{\rm BH}$--$\langle \rho \rangle_{\rm soi}$--$n_{\rm sph}$ plane. 
We shall leave this for future work. It is worth noting that exact estimates of $\Gamma_{\rm TDE}$ can vary depending on the presence of a nuclear star cluster.

\section{Conclusion}
\label{conclusions}
We used the largest-to-date sample of galaxies which have a careful multi-component decomposition of their projected surface brightness profile \citep{Savorgnan:Graham:2016:I, Davis:2018:a, Sahu:2019:a} and a directly-measured central black hole mass present in the literature (Section \ref{Data}). 
We build upon our recent (published) work, where we revealed  morphology-dependent $M_{\rm BH}$--($M_{\rm *,sph}$ and $M_{\rm *,gal}$) relations \citep{Davis:2018:b, Davis:2018:a, Sahu:2019:a},  $M_{\rm BH}$--$\sigma$ relations \citep{Sahu:2019:b}, and  $M_{\rm BH}$--($n_{\rm sph}$ and $\rm R_{e, sph}$) relations \citep{Sahu:2020}.  

Here, we investigated the connection between the black hole mass and the host spheroid's projected and internal stellar mass densities (Sections \ref{Projected_density} and \ref{Deprojected_density}, respectively).  
More specifically, we presented the scaling relations of  $M_{\rm BH}$ with the spheroid projected luminosity density ($\mu$, $\rm mag \, arcsec^{-2}$) and projected stellar mass density ($\Sigma$ and $\langle \Sigma \rangle $,  $\rm M_\odot \, pc^{-2}$) at and within 
various spheroid radii (e.g., $\rm R=0, 1\, kpc, 5\, kpc,$ and $ R_{\rm e, sph}$).

Importantly, we explored the correlation of $M_{\rm BH}$ with the internal stellar mass density $\rho$ ($\rm M_\odot \, pc^{-3}$), which is a better measure of density than the projected column density. 
We deprojected the (S\'ersic) surface brightness profiles of our galactic spheroids using the inverse Abel transformation (Appendix \ref{calculation_density}) and numerically calculated the internal densities at various internal radii, including the black hole's sphere-of-influence radius ($r_{\rm soi}$), fixed physical internal radii (e.g., 1 kpc, 5 kpc), and the spatial half-mass radius $r_{\rm e, sph}$. 
We investigated possible correlations between $M_{\rm BH}$ and the internal stellar mass density at and within these spheroid radii. We also presented the  density profiles (Figure \ref{Rho_profiles}), which help in understanding the  various observed $M_{\rm BH}$--$\rho$ correlations (Table \ref{fit parameters2}). 

In all these cases, we explored the dependence of the black hole scaling relations on the host galaxy morphology, i.e., possible division/substructure in the scaling diagrams due to ETGs versus LTGs, S\'ersic versus core-S\'ersic spheroids, barred versus non-barred galaxies, and galaxies with and without a stellar disk. The main results are summarized below.  

\begin{itemize}

\item{Spheroids with higher $M_{\rm BH}$ have a brighter central surface brightness $\mu_{\rm 0, sph}$ (Equation \ref{Mbh_mu0}) or higher central projected stellar mass density $\Sigma_{\rm 0,sph}$ (Figure \ref{Mbh_I0_Ie}). This is true for S\'ersic spheroids without depleted cores.
This is qualitatively consistent with the linear $M_{\rm BH}$--$\mu_{\rm 0, sph}$ relation predicted in \citet{Graham:Driver:2007}. However, the total rms scatter in the $M_{\rm BH}$--$\mu_{\rm 0, sph}$ (and $\Sigma_{\rm 0,sph}$) diagrams are notably high ($\rm \sim1$ dex, see the fit parameters in Table \ref{fit parameters1}).}

\item{$M_{\rm BH}$ defines a positive correlation with the average projected density, $\langle \Sigma \rangle_{\rm 1 kpc,sph}$, within the inner 1 kpc  of the host spheroid (\textit{aka} the spheroid compactness). The relation has  $\Delta_{\rm rms|BH}=$ 0.69 dex (Equation \ref{Mbh_Sigma_1kpc}), and is followed by all galaxy types (see the left-hand panel of Figure \ref{Compactness}, and Section \ref{3.2}).}

\item{$M_{\rm BH}$ has a stronger correlation with  $\langle \Sigma \rangle_{\rm R,sph}$ for $\rm R >1\, kpc$, than with the spheroid compactness $\langle \Sigma \rangle_{\rm 1 kpc,sph}$, such that the slope of the relation 
and the scatter decreases with  increasing $R$. The  total scatter starts saturating at $\sim 0.59$ dex beyond $\rm \sim5\, kpc$ (see Figure \ref{Scatter_R_Sigma}). 
 }

\item{In the $M_{\rm BH}$--$\mu_{\rm e, sph}$ and $M_{\rm BH}$--$\Sigma_{\rm e, sph}$ (and $\langle \Sigma \rangle_{\rm e, sph}$) diagrams, ETGs and LTGs (S-types) follow  different negative relations (Figure \ref{Mbh_Ie}, Table \ref{fit parameters1}). 
The negative trend is because spheroids with higher $M_{\rm BH}$ have a larger half-light radius with a lower density at/within these radii relative to that of spheroids with lower $M_{\rm BH}$. 
Further investigation reveals an offset between the E- and ES/S0-type galaxies in these diagrams, with suggestively similar slopes as that of LTGs (see Figure \ref{Mbh_Ie_E_ESS0}). 
However, the correlation coefficients are very poor, and the high scatter across these relations makes it difficult to quantify this offset correctly. Moreover, the actual distributions for the E-, ES/S0-, and S-types are expected to be curved; the predicted curves are also presented in Figures \ref{Mbh_Ie} and  \ref{Mbh_Ie_E_ESS0} (Section \ref{3.3}).
}

\item{$M_{\rm BH}$ correlates with the internal density at and within the corresponding sphere-of-influence radius ($\rho_{\rm soi, sph}$ and $\langle \rho \rangle_{\rm soi, sph}$, Figure \ref{Mbh-Rho_soi}).  
The S\'ersic and core-S\'ersic galaxies seem to define two different relations with a negative slope.  
The core-S\'ersic galaxies define a  shallower $M_{\rm BH}$--$\rho_{\rm soi, sph}$ relation with $\Delta_{\rm rms|BH}=$ 0.21 dex, whereas, the S\'ersic galaxies with $\rm n \gtrsim 1$ follow a steeper relation with $\Delta_{\rm rms|BH}=$ 0.77 dex (see Table \ref{fit parameters2}). 
This substructuring is primarily due to the range of high S\'ersic index profiles for the core-S\'ersic spheroids (see the top panel of Figure \ref{Mbh-Rho_soi} and Section \ref{4.1}). 
The data suggests an $M_{\rm BH}$--$\rho_{\rm soi, sph}$--$n_{\rm sph}$ plane, which will be the subject of future work.
}

\item{Analogous to the (projected) spheroid compactness $\langle \Sigma \rangle_{\rm 1 kpc,sph}$, we introduced the spheroid spatial compactness, $\langle \rho \rangle_{\rm 1 kpc,sph}$, which is a measure of density within a sphere of 1 kpc radius. 
The quantity  $\langle \rho \rangle_{\rm 1 kpc,sph}$  defines a positive correlation with the $M_{\rm BH}$,  which has $\Delta_{\rm rms|BH}=$ 0.75 dex (see Equation \ref{Mbh_rho_1kpc} and the left-hand panel in Figure \ref{M-D1}). 
As with $\langle \Sigma \rangle_{\rm 1 kpc,sph}$, we do not find a morphological dependence in the $M_{\rm BH}$--$\langle \rho \rangle_{\rm 1 kpc,sph}$ diagram. }

\item{Analogous to the $M_{\rm BH}$--$\langle \Sigma \rangle_{\rm R,sph}$ diagram, we find stronger correlations between  $M_{\rm BH}$ and $\langle \rho \rangle_{\rm r,sph}$ for $\rm r > 1 \rm \,kpc$. The slope of the $M_{\rm BH}$--$\langle \rho \rangle_{\rm r,sph}$ relation and the total scatter decreases with increasing internal radius $r$, where $\Delta_{\rm rms|BH}$  asymptotes at $\sim 0.6$ dex for $\rm r \gtrsim 5 kpc$ (Figure \ref{Scatter_rho_r}). The $M_{\rm BH}$--$\langle \rho \rangle_{\rm 5 kpc,sph}$ relation (Equation  \ref{Mbh_rho_5kpc}) is shown in Figure \ref{M-D1}. Given the comparable scatter in the $M_{\rm BH}$--$\langle \Sigma \rangle_{\rm R,sph}$ and $M_{\rm BH}$--$\langle \rho \rangle_{\rm r,sph}$ diagrams, both the relations seem equally good predictors of  $M_{\rm BH}$, where the density within 5 kpc is preferred over the density within 1 kpc (Section \ref{4.2}).}

\item{In the $M_{\rm BH}$--$\rho_{\rm e, int, sph}$ and $M_{\rm BH}$--$\langle \rho \rangle_{\rm e, int, sph}$ diagrams, ETGs and LTGs  appear to define two different relations with a negative slope (top panels in Figure \ref{Mbh-Rho_re}). 
Further analysis reveals that ETGs with a disk (E) and ETGs without a disk (ES/S0) appear to follow two different almost parallel relations, offset by more than an order of magnitude in the $M_{\rm BH}$-direction (bottom panels in Figure \ref{Mbh-Rho_re}). They roughly have the same slope ($\sim-1$) as the relation for LTGs (Table \ref{fit parameters2}). However, the relation may be curved, in which case the observed slope is a function of our sample's range of black hole mass.
This morphology-dependent pattern has also been seen in the $M_{\rm BH}$--$M_{\rm *,sph}$ \citep{Sahu:2019:a}, $M_{\rm BH}$--$R_{\rm e,sph}$ \citep{Sahu:2019:b}, and $M_{\rm BH}$--$\Sigma_{\rm e,sph}$ diagrams (Figures \ref{Mbh_Ie} and   \ref{Mbh_Ie_E_ESS0}). 
}

\end{itemize}


The revelation of morphology-dependent substructure in diagrams of black hole mass with various host spheroid/galaxy properties makes it more complex to conclude which relation may be the best to predict $M_{\rm BH}$ or the most fundamental relation. 
It also rewrites the notion of the coevolution of galaxies and their black holes. The black holes appear to be aware of the galaxy morphology and thus the formation physics of the galaxy.

The central densities ($\mu_{\rm 0, sph}$, $ \Sigma_{\rm 0, sph}$,  $\langle \rho \rangle_{\rm soi, sph}$, and $\rho_{\rm soi, sph}$) are based on the inward extrapolation of the  S\'ersic component of the spheroid's surface brightness model; however, additional nuclear star clusters or partially depleted cores will modify these densities. 
In future work, we hope to use high-resolution HST images to measure the depleted cores of the core-S\'ersic galaxies and extract the nuclear star clusters from the host galaxy profile. 
This will enable us to revisit the  $M_{\rm BH}$--central density relations. 

The $M_{\rm BH}$--density relations revealed in this paper have a wide range of applications (Section \ref{Discussion}). 
For example: an alternative way to estimate the black hole mass in other galaxies;  forming tests for  realistic simulated galaxies with a central black hole; estimating the SMBH binary merger time scales; constraining the orbital frequency of the SMBHB during the transition from binary hardening to the GW emission phase; modeling the tidal disruption event rates (e.g.,  Equation \ref{TDE}); estimating/modeling the SMBH binary merger rate; and modifying the characteristic strains for the detection of long-wavelength gravitational waves for pulsar timing arrays and space interferometers.

\begin{acknowledgements}
This research was conducted with the Australian Research Council Centre of Excellence for Gravitational Wave Discovery (OzGrav), through project number CE170100004. This project was supported under the Australian Research Council's funding scheme DP17012923. This material is based upon work supported by Tamkeen under the NYU Abu Dhabi Research Institute grant CAP$^3$. I additionally thank the astrophysics group at the University of Queensland for hosting me and providing an office space for one year during the covid-19 pandemic. 
\end{acknowledgements}

\appendix

\section{Calculation of The Bulge Internal Density}
\label{calculation_density}

The surface brightness (projected/column luminosity density) profile of a galactic spheroid or an elliptical galaxy is very well described using the \citep{Sersic:1963, Sersic:1968} function, which can be expressed as   
\begin{IEEEeqnarray}{rCl}
\label{Sersic}
\rm I(R)= I_e \exp{\left[-b \left \{ \left(\frac{R}{R_e} \right)^{1/n}-1 \right \}\right]}.
\end{IEEEeqnarray}
It is parametrized by the S\'ersic index ($\rm n$), the scale radius ($\rm R_e$), and the intensity ($\rm I_e$) at $\rm R_e$. 
The term $\rm b$ is a function of n, defined such that the scale radius $\rm R_{e}$  encloses 50\% of the total spheroid light; therefore, $\rm R_{e}$ is known as the (projected) effective half-light radius\footnote{See \citet{Graham:Re:2019} for a detailed review of popular galactic radii and \citet{Graham:Driver:2005} for an overview of the S\'ersic model.}. 
As noted by \citep{Ciotti:1991},  the exact value of $\rm b$ can be obtained using $\rm  \Gamma(2n)=2 \gamma(2n, b)$ or it can be approximated as $\rm b=1.9992 \, n - 0.327$ for the value of n between 0.5 to 10 \citep{Capaccioli:1989}. 
The parameter $\rm n$ is the profile \textit{shape parameter} and quantifies the central light concentration of the spheroid  \citep{Trujillo:Graham:Caon:2001}. 
The intensity, $\rm I_e$,  is related to the surface brightness ($\mu \rm \, in \, mag/arcsec^{2}$) at $\rm R_{e}$, via $\rm \mu_e  \equiv  -2.5\log(I_e)$. 

As mentioned in Section \ref{Data}, the bulge S\'ersic profile parameters used here are taken from \citet[][their table A1]{Sahu:2020}. 
Who provide both major-axis bulge surface brightness parameters ($\rm n_{maj}, R_{e, maj}, \mu_{e, maj}$) and the equivalent-axis bulge surface brightness parameters ($\rm n_{eq}, R_{e, eq}, \mu_{e, eq}$) obtained by  independent multi-component decompositions of galaxy surface brightness profiles along the major-axis and geometric mean-axis (equivalent to a circularised axis), respectively, \citep[see][for more details on the decomposition process]{Sahu:2019:a}. 
As described below, we used the equivalent-axis bulge parameters to utilize their circular symmetry while calculating the (deprojected) internal bulge density profile.

Using the inverse Abel (integral) transformation  \citep{Abel1826, Anderssen1990}, the spatial luminosity density, $j(r)$, for a spherical system can be expressed in terms of derivative ($\rm d \, I(R)/d \, R$) of the projected luminosity density profile \citep[see][]{BinneyTremaine1987}  as, 
\begin{IEEEeqnarray}{rCl}
\label{Abel}
\rm j(r)= -\frac{1}{\pi} \int_{r}^{\infty} \frac{d \rm I(R)}{d \rm R} \frac{d \rm R}{\sqrt{\rm R^2 -\rm r^2}},
\end{IEEEeqnarray}
where R represents a projected radius and r represents a 3D spatial (or internal) radius. Using the S\'ersic profile (Equation \ref{Sersic}) for $\rm I(R)$ and a stellar mass-to-light ratio, $\Upsilon_\lambda$ (suitable for the corresponding image wavelength $\lambda$), to convert the luminosity into stellar mass, the spatial mass density profile, $\rho(r) \equiv  \Upsilon_\lambda \,j(r)$, can be expressed in a simplified integral form \citep{Ciotti:1991, GrahamColless1997} as, 
\begin{IEEEeqnarray}{rCl}
\label{Rho_int}
\rm \rho(r) = \Upsilon_\lambda \frac{\rm I_e b e^{b}}{\pi r} \left(\frac{r}{R_e} \right)^{1/n} \int_{0}^{1} \frac{e^{(-b (\rm r/R_e)^{1/n}/t)}}{\rm t^2 \sqrt{\rm t^{-2n} -\rm 1}}d\rm t.
\end{IEEEeqnarray}
The above transformation (Equation \ref{Abel}) is applicable for a spherical system, and we can use the equivalent-axis bulge S\'ersic parameters ($\rm n_{eq}$, $\rm R_{e, eq}$, and $\rm I_{e, eq}$) to obtain $\rm \rho(r)$. 

Using $\rm R_e$ in parsec (pc), $\rm I_{e}$ in solar luminosity per unit area ($ \rm L_{\odot}/pc^2$), and $\Upsilon_\lambda$ in the units of solar mass per solar luminosity ($\rm M_\odot / L_\odot$), we obtain $\rm \rho(r)$ in the units of $\rm M_{\odot} /pc^3$ from the above integral (Equation \ref{Rho_int}). The surface brightness $\mu_e$ (in $\rm mag/arcsec^2$) at the half-light radius can be mapped into $\rm I_{e}$ ($ \rm L_{\odot}/pc^2$) using the following equation taken from \citet{Graham:Merritt:2006},
\begin{IEEEeqnarray}{rCl}
\label{Ie}
\rm -2.5 \log(I_e \, \rm [L_\odot pc^{-2}]) &=& \rm  \mu_e - DM - \mathfrak{M}_{\odot, \lambda}  -  2.5 \log(1/s^2),
\end{IEEEeqnarray}
where $\rm DM=25+5\log[ Distance\, (Mpc) ]$ is the distance modulus, $\rm \mathfrak{M}_{\odot, \lambda}$ is the absolute magnitude of the Sun in the corresponding wavelength-band $\lambda$, and $s$ is the physical size scale for a galaxy in $\rm pc \, arcsec^{-1}$. The projected (or surface) mass density $\Sigma \, \rm (M_\odot \,pc^{-2})$ at any projected radius R is calculated via,
\begin{IEEEeqnarray}{rCl}
\label{Sigma_R}
\rm -2.5 \log(\Sigma_R \, \rm [M_\odot pc^{-2}]) &=& \rm  \mu_R - DM - \mathfrak{M}_{\odot, \lambda}  -  2.5 \log(1/s^2)-2.5 \log(\Upsilon_\lambda).
\end{IEEEeqnarray} 
The solution of the above integral (Equation \ref{Rho_int}) can be expressed with the Meijer-G function\footnote{For some cases, it turns out as a sum of generalized hypergeometric function residues. See \url{http://functions.wolfram.com/HypergeometricFunctions/MeijerG/26/01/02/} for how Meijer-G function and Hypergeometric functions are linked.} \citep{Meijer1936, Bateman1953, MazureCapelato2002}, which we numerically calculated using a \textit{Mathematica} script to obtain the internal densities at various spheroid radii, used in Section \ref{Deprojected_density}. 

\citet[][PS hereafter]{Prugniel1997} provides a remarkably simple, and one of the closest, approximation to the deprojected S\'ersic profile (Equation \ref{Rho_int}), which can be expressed as,
\begin{IEEEeqnarray}{rCl}
\label{PS}
\rm \rho(r)= \rho_{\rm e} \left(\frac{r}{R_e} \right)^{-p} \exp \left\lbrace -b \left[  \left(\frac{r}{R_e} \right)^{1/n} -1 \right] \right\rbrace.
\end{IEEEeqnarray}
Here, $\rho_{\rm e}$ is the spatial mass density at $\rm r= R_e$, and $p$ is a function of $n$, obtained by maximizing the agreement between this approximated $\rho(r)$ profile (Equation \ref{PS}) and the exactly deprojected (S\'ersic) density profile (Equation \ref{Rho_int}).  
The value of $p$ is given by $p= 1.0 - 0.6097/n + 0.05563/n^2 $ for a radial range of $ 10^{-2} \lesssim  r/R_e \lesssim 10^{3}$ and index range of $ 0.6 \lesssim n \lesssim 10 $ \citep{LimaNeto:1999,Marquez:LimaNeto:2000}. On equating the total mass obtained from the projected S\'ersic profile (Equation \ref{Sersic}) with the total mass calculated using the \citetalias{Prugniel1997} spatial density profile (Equation \ref{PS}), considering a constant   mass-to-light ratio, one has
\begin{IEEEeqnarray}{rCl}
\label{Rhoe_Ie}
\rm \rho_{\rm e} =  \Upsilon \left(\frac{I_e}{2 R_e b^{n(p-1)}}\right)   \left[  \frac{\Gamma(2n)}{\Gamma(n(3-p)) }  \right]. 
\end{IEEEeqnarray}

Owing to its simple analytical form and the model parameters common to the S\'ersic  luminosity profile,  the \citetalias{Prugniel1997} model makes it easy to estimate the internal density profile of elliptical galaxies and the spheroids of multi-component (i.e., ES-, S0-, and Spiral-type) galaxies. Thus, Equations \ref{Rho_int} and \ref{PS}, both, are applicable for a galaxy/component whose surface brightness profile can be described using a  S\'ersic function; however,  Equation \ref{Rho_int} can provide the most accurate value.

For  core-S\'ersic galaxies, i.e., galaxies with power-law $+$ S\'ersic spheroid surface brightness profiles, \citet[][their equation 5]{Terzic:Graham:2005} modified the \citetalias{Prugniel1997} model and presented an expression for the deprojected core-S\'ersic spheroid density profile. 
However, as we do not have precise parameters for the power-law core of our core-S\'ersic galaxies, we use the S\'ersic part of their surface brightness profile and deproject its inward extrapolation to obtain the central/inner $rho$ for core-S\'ersic galaxies.

\begin{figure}
\begin{center}
\includegraphics[clip=true,trim= 00mm 00mm 15mm 15mm,width=   0.5\columnwidth]{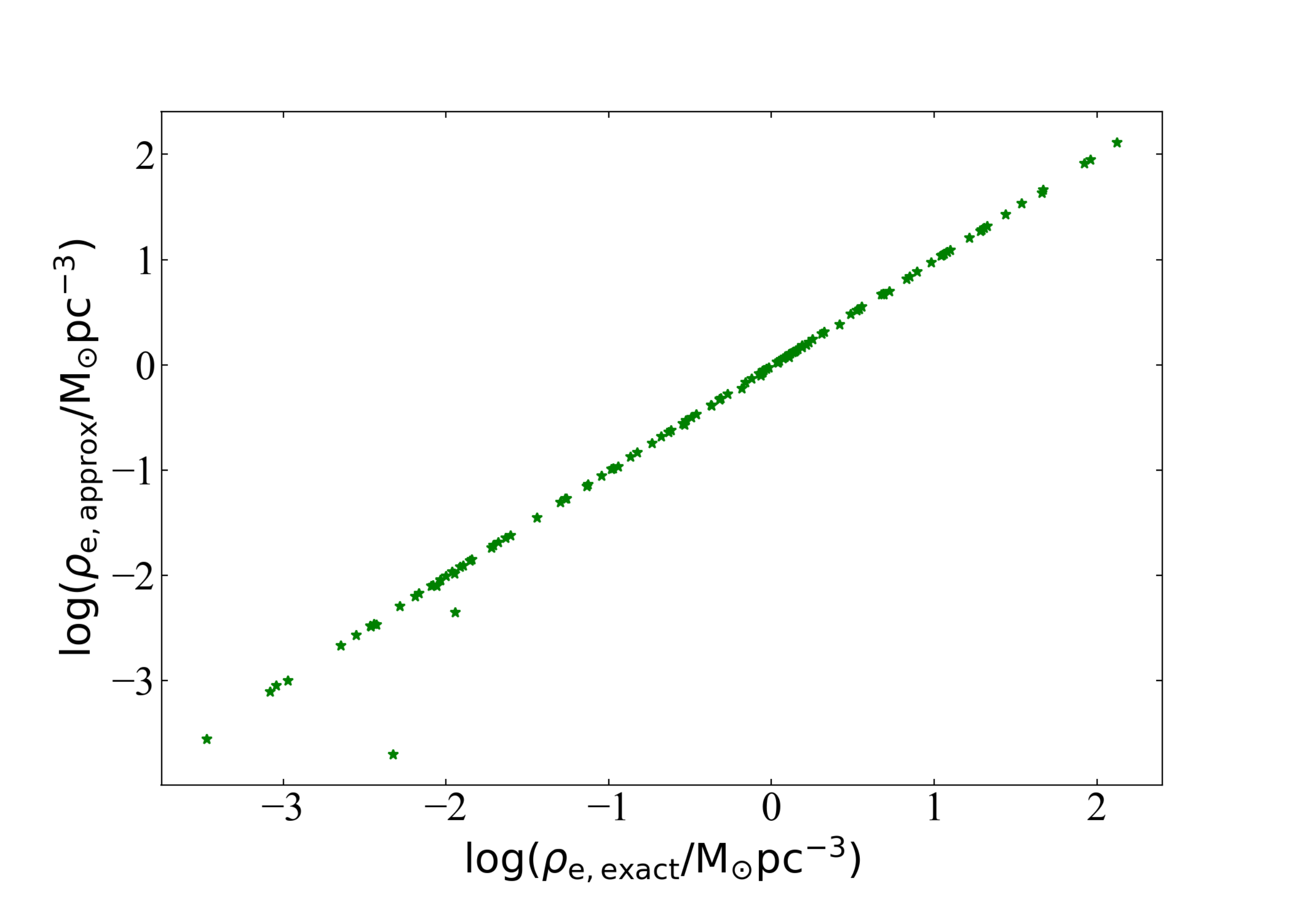}
\caption{The spatial density at $\rm R_e$ calculated using the \citetalias{Prugniel1997}  model ($\rho_{\rm e, approx}$) plotted against the numerically calculated (Equation \ref{Rho_int}) spatial density ($\rho_{\rm e, exact}$).
}
\label{Rhoe-Rhoe}
\end{center}
\end{figure} 

The approximation of the deprojected density profiles can be imprecise ($\rm <10\% \, diferrence \, at \, 0.01< R/R_e <100\, R_e \, for \, n>2$) to emulate the actual density profiles at the central radii, especially for low S\'ersic index spheroids. See the comparisons in  \citet[][their figure 4]{Terzic:Graham:2005}, \citet{Emsellem2008}, and  \citet{Vitral2020}. Therefore, for our black hole--internal density correlations, we prefer to use the numerically calculated internal densities from Equation \ref{Rho_int}. 

In Figure \ref{Rhoe-Rhoe}, we have compared $\rho_{\rm e, approx} $ at $r= R_{\rm e}$ calculated using the \citetalias{Prugniel1997}  model (Equation \ref{Rhoe_Ie}) against $\rho_{\rm e, exact}$, numerically calculated using Equation \ref{Rho_int}. 
Here, we see an almost one-to-one match between the two values, except for galaxies M~59, NGC~1399, and NGC~3377, the three offset galaxies in Figure \ref{Mbh_I0_Ie} with $n_{\rm sph,eq} \gtrsim 8.8$.  
The two offset points shown in Figure \ref{Rhoe-Rhoe} are  M~59 and NGC~1399, whereas, for NGC~3377, the exact integral (Equation \ref{Rho_int}) did not converge to provide an appropriate value of $\rho_{\rm e, exact}$. 


Given the agreement between the exact and approximate internal densities at $\rm r=R_e$ for the majority of the sample, for NGC~3377, we have used $\rho_{\rm e}$ obtained from the \citetalias{Prugniel1997} model. 
Similarly, for some instances, where the exact $\rho(r)$ integral (Equation \ref{Rho_int}) did not converge or provide a valid density value, we used the internal densities obtained using the  \citetalias{Prugniel1997} model (Equation \ref{Rhoe_Ie}).  
This does not have a significant effect on the best-fit relations presented here. The extended density profiles in Figure \ref{Rho_profiles} are obtained using the \citetalias{Prugniel1997} model, as it can still explain the qualitative nature of the trends observed in the $M_{\rm BH}$--$\rho$ diagrams.


\bibliography{bibliography}

\end{document}